\documentclass[aps,prd,11pt,superscriptaddress,amssymb,amsmath,nofootinbib]{revtex4-1}

\usepackage{hhline}
\usepackage{color}
\usepackage{epsfig}
\usepackage{graphicx}
\usepackage{float}
\usepackage{epstopdf}
\usepackage{hyperref}
\usepackage[usenames,dvipsnames]{xcolor}
\usepackage{multirow,booktabs}
\usepackage[normalem]{ulem}
\usepackage[d]{esvect}
\usepackage[misc,geometry]{ifsym}
\usepackage{harpoon}

\begin{document}

 \title{The geometric correlations of leptonic mixing parameters}
 \author{Ding-Hui Xu}\email{3036602895@qq.com}
 \author{Shu-Jun Rong\Letter}\email{rongshj@glut.edu.cn}
 \affiliation{College of Science, Guilin University of Technology, Guilin, Guangxi 541004, China}

 \begin{abstract}
Leptonic mixing patterns are usually extracted on the basis of groups or algebraic structures.
In this paper, we introduce an alternative geometric method to study the correlations between the leptonic mixing parameters.
At the 3$\sigma$ level of the recent global fit data of neutrino oscillations, the distribution of the scattered points of the angles between the vectors, which are constructed by the element of the leptonic mixing matrix, is analysed.
We find that the scattered points are concentrated on several special regions.
Using the data in these regions, correlations of the leptonic mixing angles and the Dirac CP violating phase are
obtained.
The implications of the correlations are shown through the predicted flavor ratio of high-energy astrophysical neutrinos (HANs) at Earth.
\end{abstract}

 \maketitle

\section{Introduction}
The discovery of neutrino flavor oscillation is a tremendous achievement in particle physics during the last two decades\cite{1,2,3}, which manifests that neutrinos have masses, and the leptonic mixing matrix is non-trivial. However, the origin of the flavor mixing is elusive for theorists.
Before the determination of the reactor mixing angle $\theta_{13}$, there are various candidates of the mixing patterns such as bi-maximal mixing (BM)\cite{4}, tri-bi-maximal mixing (TBM)\cite{5,6}, golden ratio mixing (GRM)\cite{7,8,9} and hexagonal mixing (HM)\cite{10}, etc.\cite{11,12}, for reviews see\cite{13,14}.
These patterns, especially the TBM, derived from flavor groups such as $A_4$\cite{15,16} and $S_4$\cite{16,17}, are compatible with previous published data of atmospheric neutrino and solar neutrino.
The discovery of a relatively large reactor mixing angle $\theta_{13}$\cite{3,18,19,20,21} and the remarkable progress of measurements of neutrino oscillation parameters, however, strictly constrain or exclude the above patterns.
Hence, diverse methods to generate a non-zero reactor mixing angle $\theta_{13}$ have been proposed.
A popular approach is adding a model-independent perturbation to the leading TBM matrix\cite{22}.
Another method introduces a generalized CP transform on the basis of a discrete flavor symmetry, i.e., a finite flavor group $G_{f}$ and a CP symmetry are combined\cite{23,24,25,26,27}.
Furthermore, a novel mathematical structure called group algebra was introduced\cite{28,29}. In this scenario, a specific leptonic mixing pattern could correspond to a set of equivalent elements of a group algebra.

In this paper, we introduce an alternative geometric method to extract the correlations between the leptonic mixing parameters.
This method is developed from the well-known $\mu-\tau$ reflection symmetry\cite{30,31,32,33,34,35,36,37,38,39} which denotes that two row vectors $\overrightarrow{\mu}$=($|U_{\mu1}|^2, |U_{\mu2}|^2, |U_{\mu3}|^2$) and $\overrightarrow{\tau}$=($|U_{\tau1}|^2, |U_{\tau2}|^2, |U_{\tau3}|^2$) are identical.
Here $U_{\alpha i}$($\alpha = \mu, \tau$, and i=1, 2, 3) is the element of the Pontecorvo-Maki-Nakagawa-Sakata(PMNS) mixing matrix\cite{40,41,42}
\begin{equation}
\label{eq:1}
U=
\left(
    \begin{array}{ccc}
    c_{12}c_{13}&s_{12}c_{13}&s_{13}e^{-i\delta}\\
    -s_{12}c_{23} -c_{12}s_{13}s_{23}e^{i\delta}&c_{12}c_{23}-s_{12}s_{13}s_{23}e^{i\delta}&c_{13}s_{23}\\
    s_{12}s_{23}-c_{12}s_{13}c_{23}e^{i\delta}&-c_{12}s_{23}- s_{12}s_{13}c_{23}e^{i\delta}&c_{13}c_{23}\\
    \end{array}
  \right),
\end{equation}
where $s_{ij}\equiv\sin\theta_{ij}$(for ij = 12, 13, 23), $c_{ij}\equiv\cos\theta_{ij}$ and $\delta$ is the Dirac CP phase.
Throughout this article we do not consider the Majorana phases because they are irrelevant to neutrino oscillations.
As is known, for the realistic mixing matrix, the $\mu-\tau$ reflection symmetry predicts that $\theta_{23}$=$\pi$/4, $\delta$=$\pm\pi/2$.
According to the recent global analysis NuFIT 5.0\cite{43}, the maximal CP phase is favored by the $1\sigma$ allowed region in the case of inverted mass ordering (IO) while disfavored by the  $1\sigma$ allowed region in the normal ordering (NO) case. Therefore, the $\mu-\tau$ reflection symmetry may need modifications in the near future when a precise measurement of $\delta$ would be available.

Geometrically speaking, the $\mu-\tau$ reflection symmetry means that the angle between the 3-dimensional real vectors $\overrightarrow{\mu}$ and $\overrightarrow{\tau}$ is zero.
A perturbed $\mu-\tau$ symmetry corresponds to a small but non-zero included angle. Thus, the leptonic mixing pattern and the breaking of $\mu-\tau$ reflection symmetry can be represented by the geometric quantities, namely the angles
between the row vectors $\overrightarrow{\mu}$, $\overrightarrow{\tau}$, and $\overrightarrow{e}$=($|U_{e1}|^2, |U_{e2}|^2, |U_{e3}|^2$).
In order to extract promising mixing patterns, we scan the included angles between the row vectors at the 3$\sigma$ level of the global fit data\cite{43}.
We find that the scattered points of the angles are concentrated on several special areas which are stable under the random takings of the mixing parameters at the 3$\sigma$ level.
Furthermore, it is not surprising that the approximated $\mu-\tau$ reflection symmetry corresponds to one of the special regions in the case of IO.
Therefore, it is plausible that the promising correlations of the mixing parameters are indicated in these dense regions.
On the basis of this assumption, the leptonic mixing patterns are read out from the dense zones and their implications are examined by a specific application, namely the predicted flavor ratio of high-energy astrophysical neutrinos (HANs) at Earth.

The paper is organized as follows. In Sec. \uppercase\expandafter{\romannumeral2}, we show the definition of included angles between the row vectors $\overrightarrow{e}$, $\overrightarrow{\mu}$ and $\overrightarrow{\tau}$ and extract correlations of the angles at the 3$\sigma$ level of the global fit data\cite{43}. On the basis of the correlations of the angles, the correlations of the leptonic mixing parameters are obtained.
In Sec. \uppercase\expandafter{\romannumeral3}, we apply these leptonic mixing parameters constrained by the geometric correlations to predict the flavor ratio of HANs at Earth. Finally, we summarize our main results.

\section{Geometric correlations at the 3$\sigma$ level of the global fit data.}
In this section, we first give the definition of the angles between the row vectors, the selection rule of typical data points in the dense zones of the scattered plots, and then present correlations between the angles, on the basis of which we derive the correlations between the leptonic mixing parameters.

\subsection{Definition of geometric correlations }
\begin{figure}
  \centering
  \includegraphics[width=.42\textwidth,height=0.21\textheight]{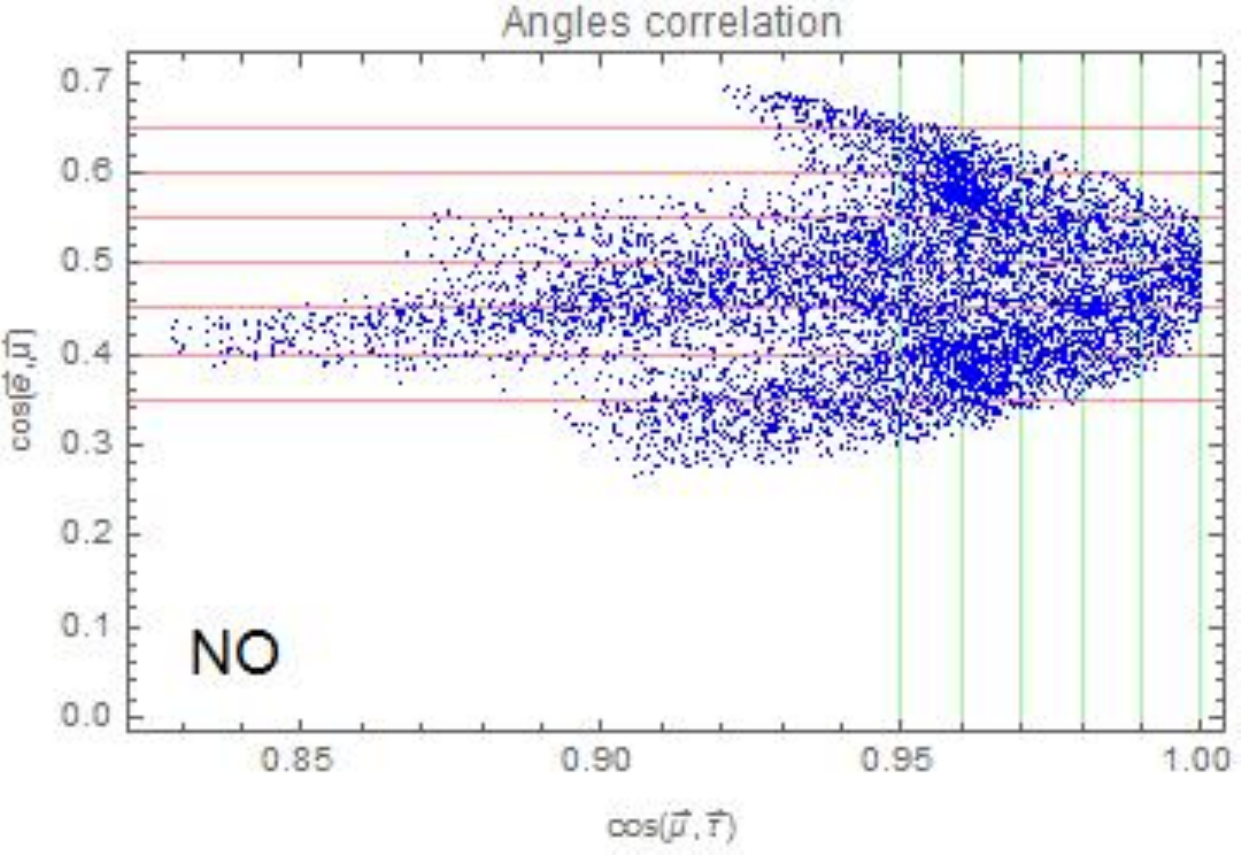}
  \hfill
  \includegraphics[width=.42\textwidth,height=0.21\textheight]{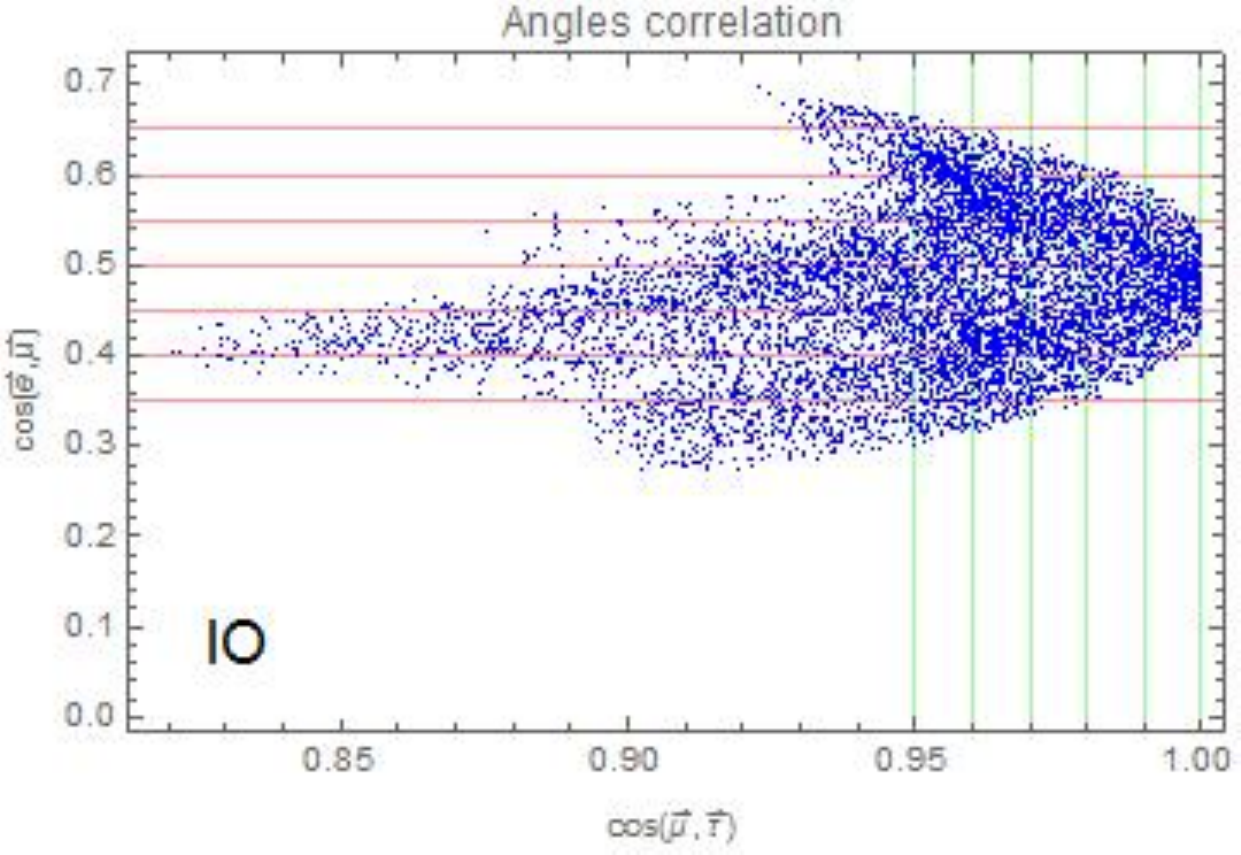}
  \hfill
  \includegraphics[width=.42\textwidth,height=0.21\textheight]{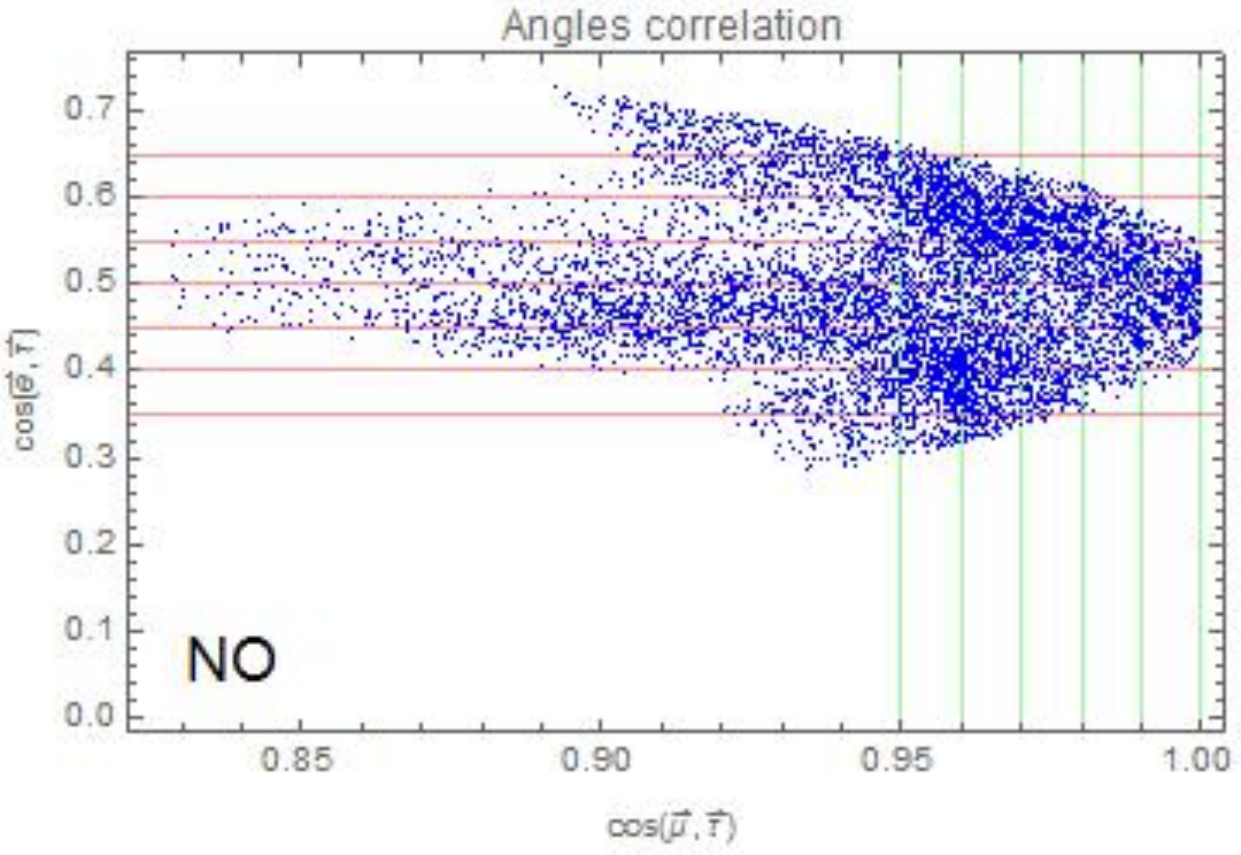}
  \hfill
  \includegraphics[width=.42\textwidth,height=0.21\textheight]{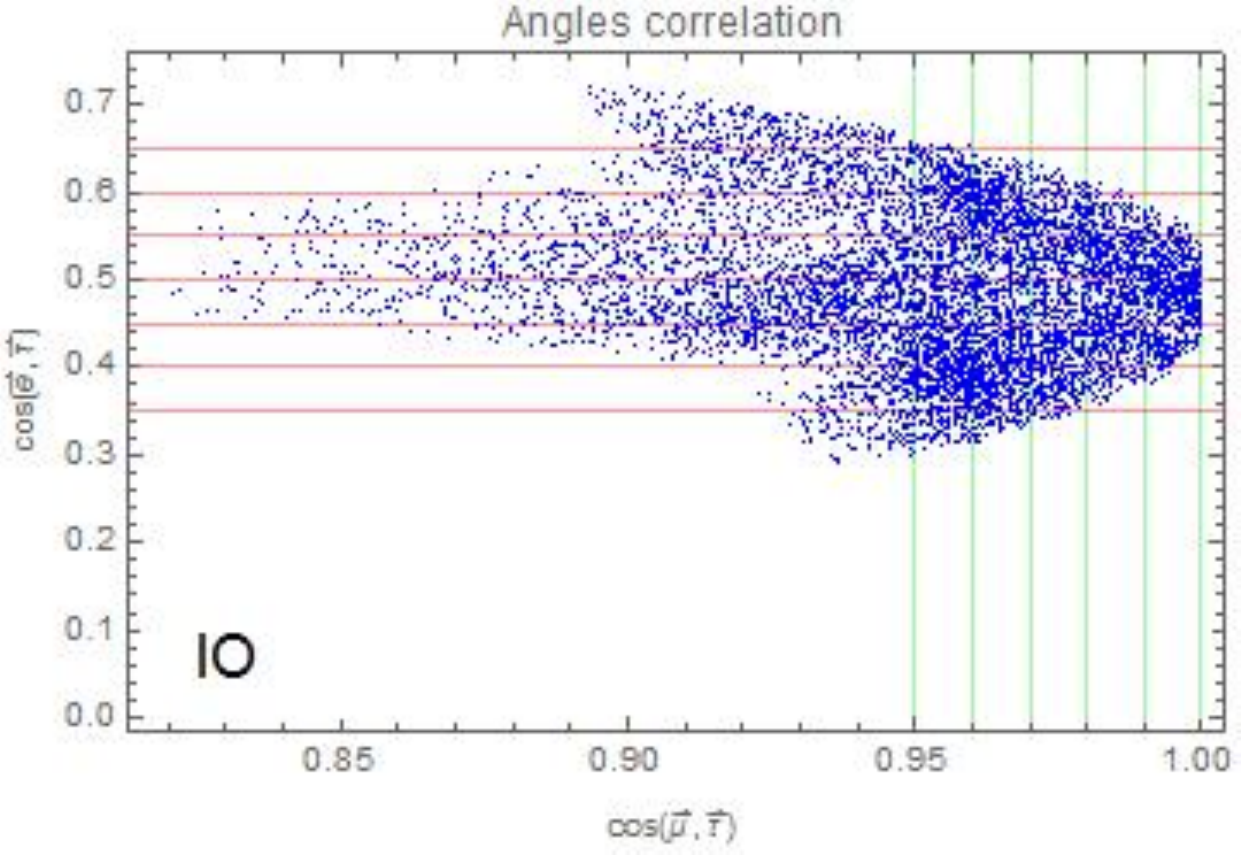}
  \caption{\label{fig:2} The front view and top view of Fig.\ref{fig:1}. Left panel: NO case, Right panel: IO case.}
\end{figure}

The included angles between the row vectors are defined as follows,
\begin{equation}
\label{eq:2}
\cos(\overrightarrow{\mu},\overrightarrow{\tau})\equiv\frac{\overrightarrow{\mu}\cdot\overrightarrow{\tau}}{|\overrightarrow{\mu}|\cdot|\overrightarrow{\tau}|}=\frac{|U_{\mu1}|^2|U_{\tau1}|^2+|U_{\mu2}|^2|U_{\tau2}|^2+|U_{\mu3}|^2|U_{\tau3}|^2}{\sqrt{|U_{\mu1}|^4+|U_{\mu2}|^4+|U_{\mu3}|^4}\cdot\sqrt{|U_{\tau1}|^4+|U_{\tau2}|^4+|U_{\tau3}|^4}},
\end{equation}
\begin{equation}
\label{eq:3}
\cos(\overrightarrow{e},\overrightarrow{\mu})\equiv\frac{\overrightarrow{e}\cdot\overrightarrow{\mu}}{|\overrightarrow{e}|\cdot|\overrightarrow{\mu}|}=\frac{|U_{e1}|^2|U_{\mu1}|^2+|U_{e2}|^2|U_{\mu2}|^2+|U_{e3}|^2|U_{\mu3}|^2}{\sqrt{|U_{e1}|^4+|U_{e2}|^4+|U_{e3}|^4}\cdot\sqrt{|U_{\mu1}|^4+|U_{\mu2}|^4+|U_{\mu3}|^4}},
\end{equation}
\begin{equation}
\label{eq:4}
\cos(\overrightarrow{e},\overrightarrow{\tau})\equiv\frac{\overrightarrow{e}\cdot\overrightarrow{\tau}}{|\overrightarrow{e}|\cdot|\overrightarrow{\tau}|}=\frac{|U_{e1}|^2|U_{\tau1}|^2+|U_{e2}|^2|U_{\tau2}|^2+|U_{e3}|^2|U_{\tau3}|^2}{\sqrt{|U_{e1}|^4+|U_{e2}|^4+|U_{e3}|^4}\cdot\sqrt{|U_{\tau1}|^4+|U_{\tau2}|^4+|U_{\tau3}|^4}}.
\end{equation}
Using the latest global fit data of the leptonic mixing parameters listed in Tab.\ref{tab:1}, the 3$\sigma$ allowed ranges of the included angle between the row vectors are shown in Fig.\ref{fig:1}.
\begin{table}
\label{tab:1}
\caption{\label{tab:1} The 3$\sigma$ allowed ranges of the leptonic mixing parameters in the global analysis NuFIT 5.0\cite{43}.}
  \centering
  \begin{tabular}{c c c}
     \noalign{\smallskip}\hline
     \noalign{\smallskip}\hline
     ~~Parameters ~~~~&~~~~ Normal ordering (NO) ~~~~&~~~~ Inverted ordering (IO)~~ \\
     \noalign{\smallskip}\hline
     ~~$\sin^2\theta_{12}$ ~~~~&~~~~ 0.269$\rightarrow$0.343 ~~~~&~~~~ 0.269$\rightarrow$0.343~~ \\
     ~~$\sin^2\theta_{13}$ ~~~~&~~~~ 0.02034$\rightarrow$0.02430 ~~~~&~~~~ 0.02053$\rightarrow$0.02436~~ \\
     ~~$\sin^2\theta_{23}$ ~~~~&~~~~ 0.407$\rightarrow$0.618 ~~~~&~~~~ 0.411$\rightarrow$0.621~~ \\
     ~~$\delta/^{\circ}$  ~~~~&~~~~ 107$\rightarrow$403 ~~~~&~~~~ 192$\rightarrow$360~~ \\
     \hline
   \end{tabular}
\end{table}
\begin{figure}
\label{fig:1}
  \centering
  \includegraphics[width=.42\textwidth,height=0.21\textheight]{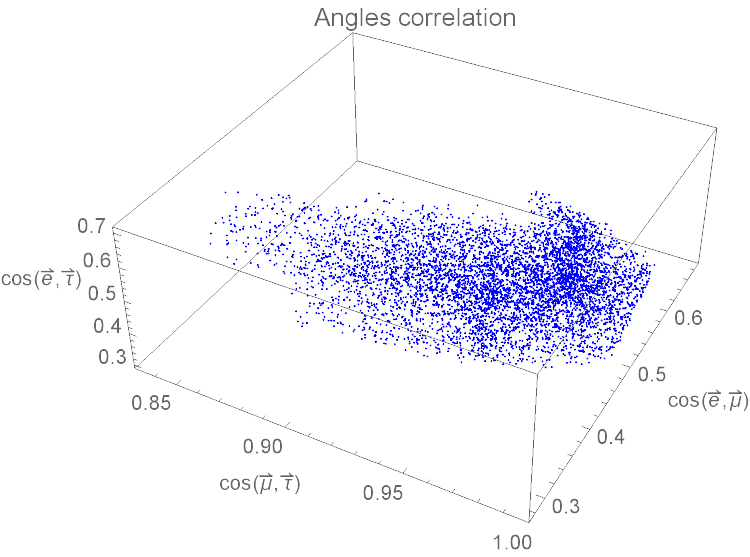}
  \hfill
  \includegraphics[width=.42\textwidth,height=0.21\textheight]{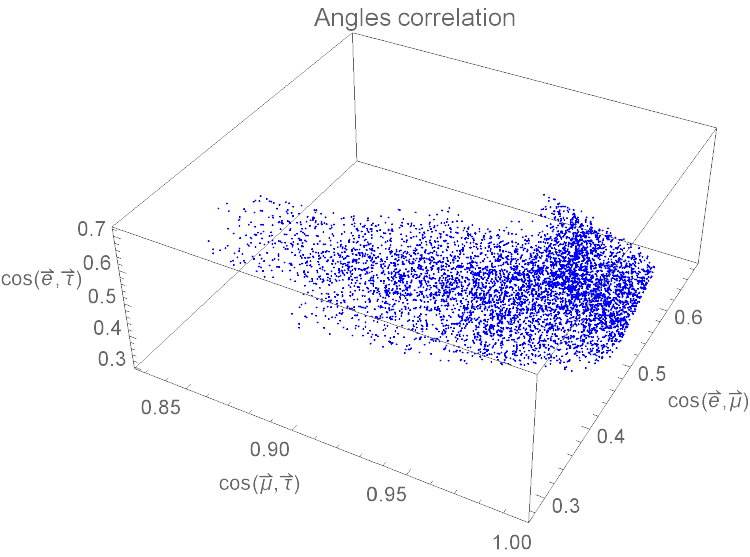}
  \caption{\label{fig:1} Three dimensional scattered plots of the included angles based on the recent global fit data listed in Tab.\ref{tab:1}. Left panel: NO case, Right panel: IO case.}
\end{figure}
We can see that the data points in these plots are of a nonuniform distribution.
There are several special regions on which the points are concentrated.
This observation is more obvious in the front view and top view of the plots shown in Fig.\ref{fig:2}.
From the 2-dimensional scattered plots, we obtain the observations as follows.\\
(a) NO case: The data points of $(\cos(\overrightarrow{\mu},\overrightarrow{\tau})$, $\cos(\overrightarrow{e},\overrightarrow{\mu}))$ are concentrated on
two regions around the points (0.96, 0.40) and (0.96, 0.59), respectively.
The data points of $(\cos(\overrightarrow{\mu},\overrightarrow{\tau})$, $\cos(\overrightarrow{e},\overrightarrow{\tau}))$ are also converged on two regions around the points (0.96, 0.39) and (0.96, 0.58), respectively. \\
(b) IO case: The data points of $(\cos(\overrightarrow{\mu},\overrightarrow{\tau})$, $\cos(\overrightarrow{e},\overrightarrow{\mu}))$ are concentrated on
three regions around the points (0.96, 0.41) , (0.96, 0.60) and (1.00, 0.49), respectively.
The data points of $(\cos(\overrightarrow{\mu},\overrightarrow{\tau})$, $\cos(\overrightarrow{e},\overrightarrow{\tau}))$ are converged on three regions around the points (0.96, 0.39), (0.96, 0.60) and (1.00, 0.48), respectively.
As is known, $\cos(\overrightarrow{\mu},\overrightarrow{\tau})$ $\sim 1$ corresponds to the approximated $\mu-\tau$ reflection symmetry.

Furthermore, we note that these dense zones of the scattered plots are stable under the random takings of the global fit data at $3\sigma$ level.
Therefore, we consider that the global fit data indicate correlations between the included angles of the row vectors.
Correspondingly, besides the leptonic mixing pattern from the approximated $\mu-\tau$ reflection symmetry, other promising patterns or correlations of the mixing parameters can be read out from the dense regions in the scattered plots.

\subsection{Selection rule of typical data points in the dense regions of the scattered plots}
In order to effectively filter out some representative data points in the dense zones of Fig.\ref{fig:2}, we divide the data samples of the angles between the row vectors into small enough intervals.
The advantage of this method is that the difference between the impacts of two arbitrary data points in the same interval on the correlations of the lepton mixing parameters can be ignored. In other words, we can use the midpoint to represent the whole interval. The selection rule is explained in more detail as follows.

\begin{figure}
  \centering
  \includegraphics[width=.49\textwidth]{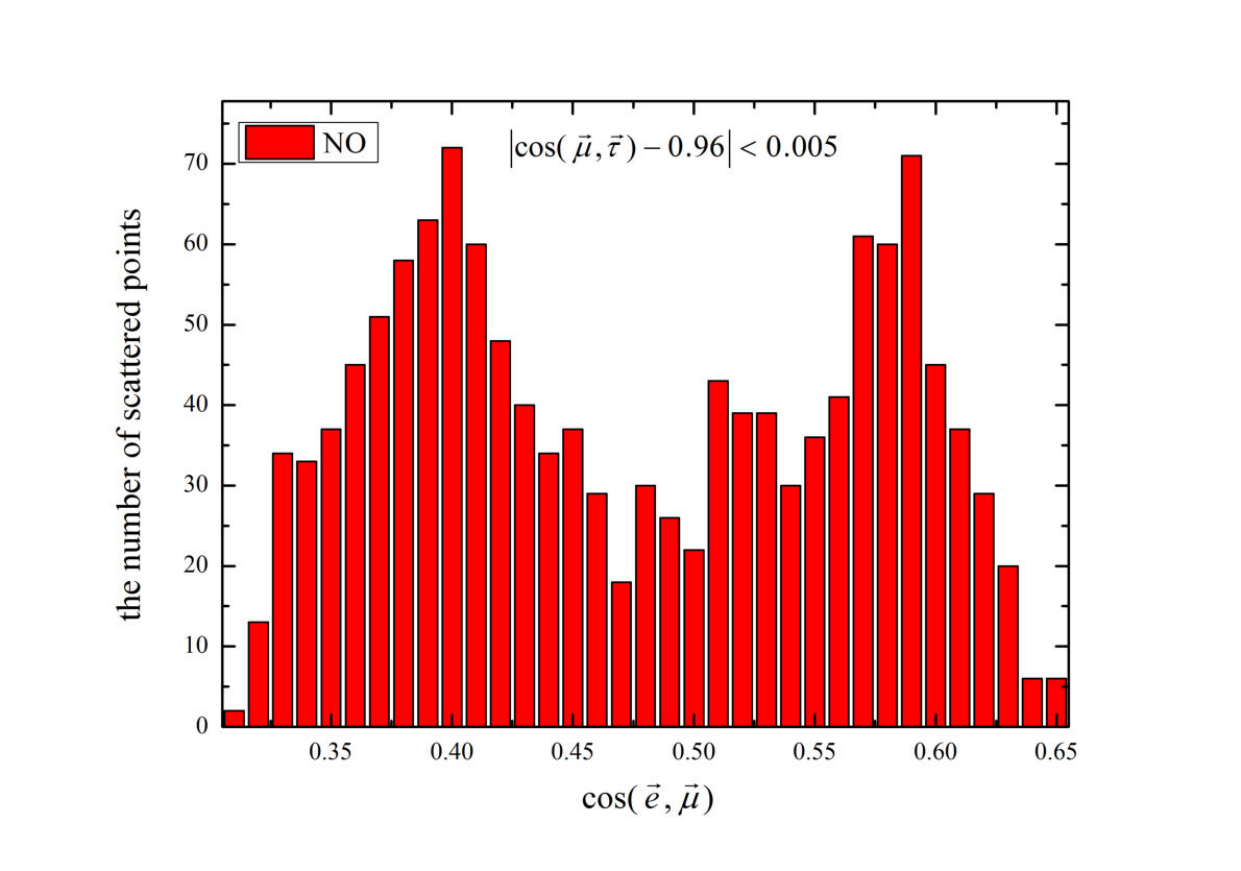}
  \hfill
  \includegraphics[width=.49\textwidth]{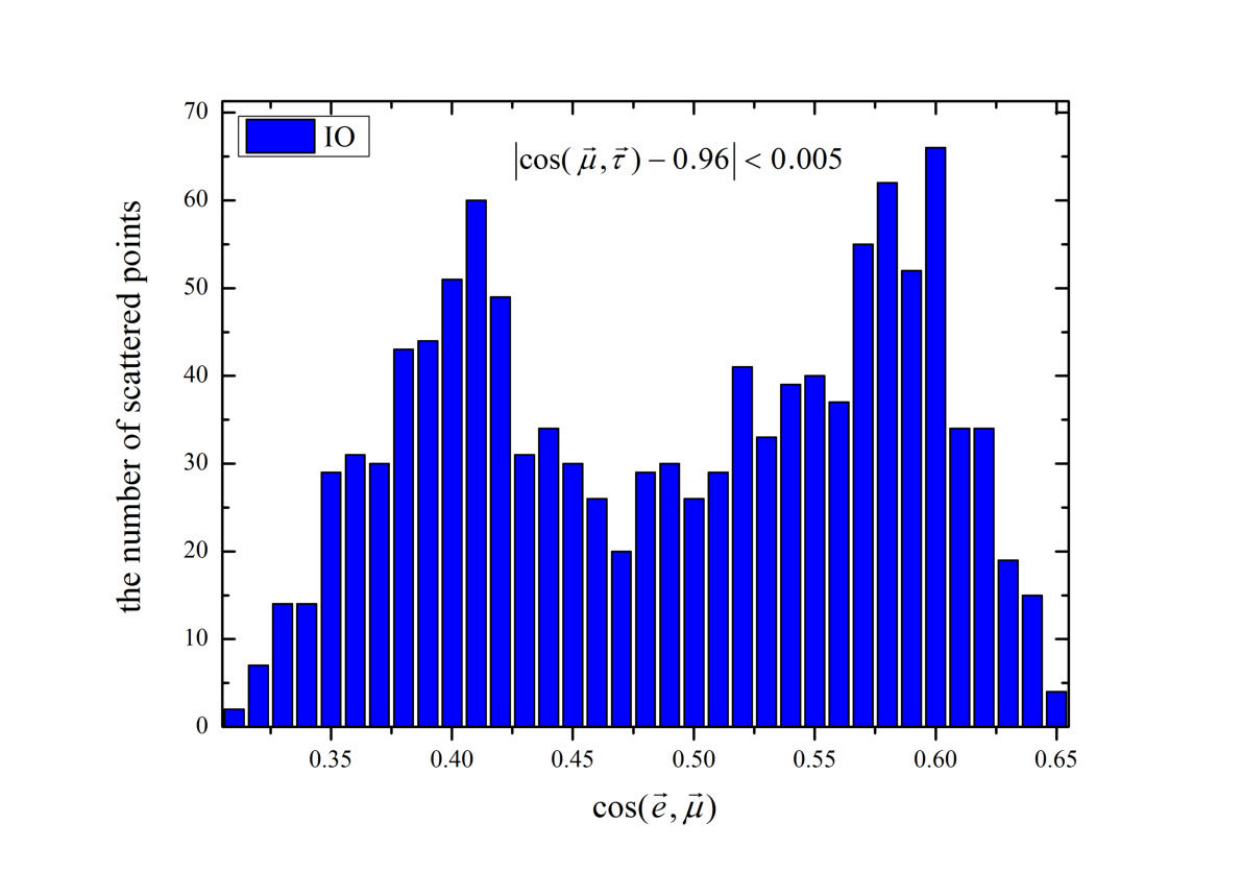}
  \hfill
  \includegraphics[width=.49\textwidth]{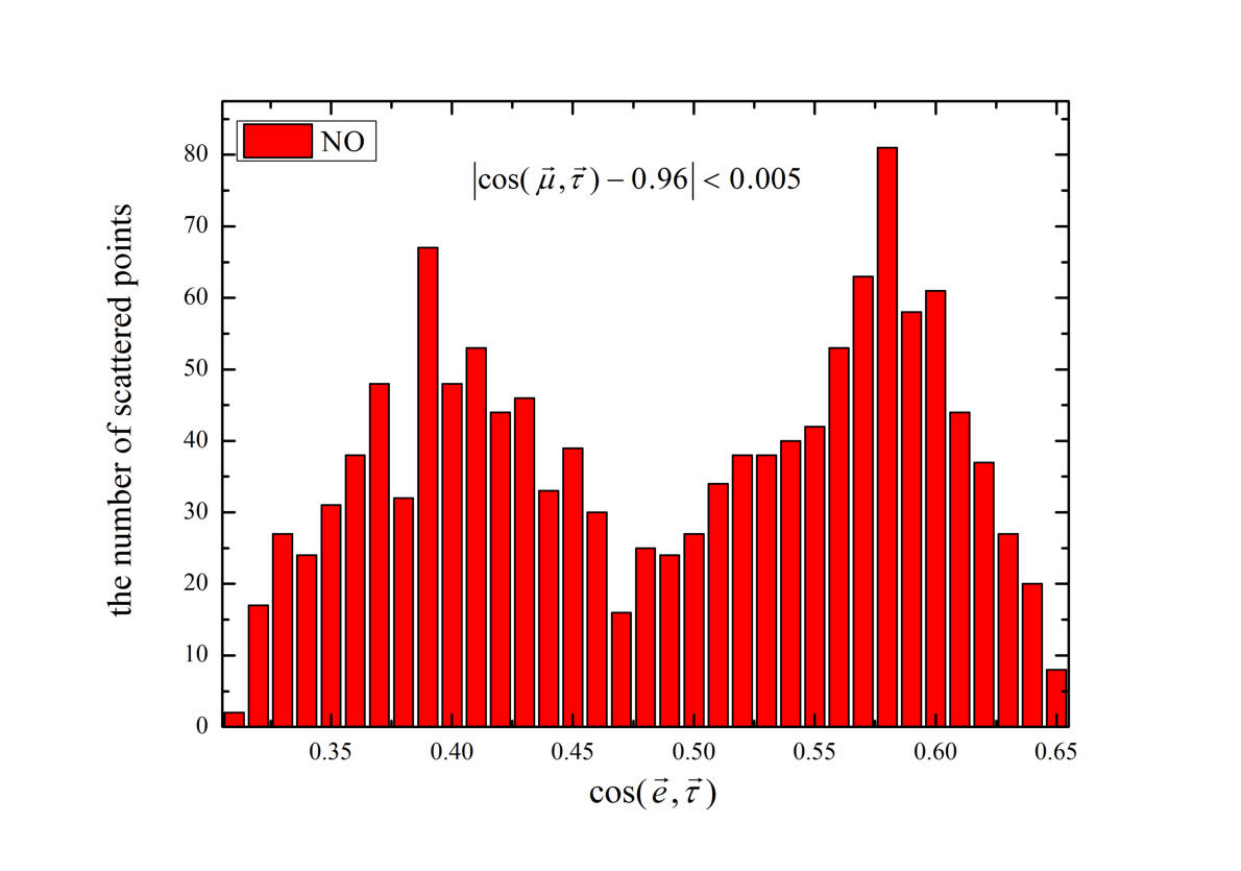}
  \hfill
  \includegraphics[width=.49\textwidth]{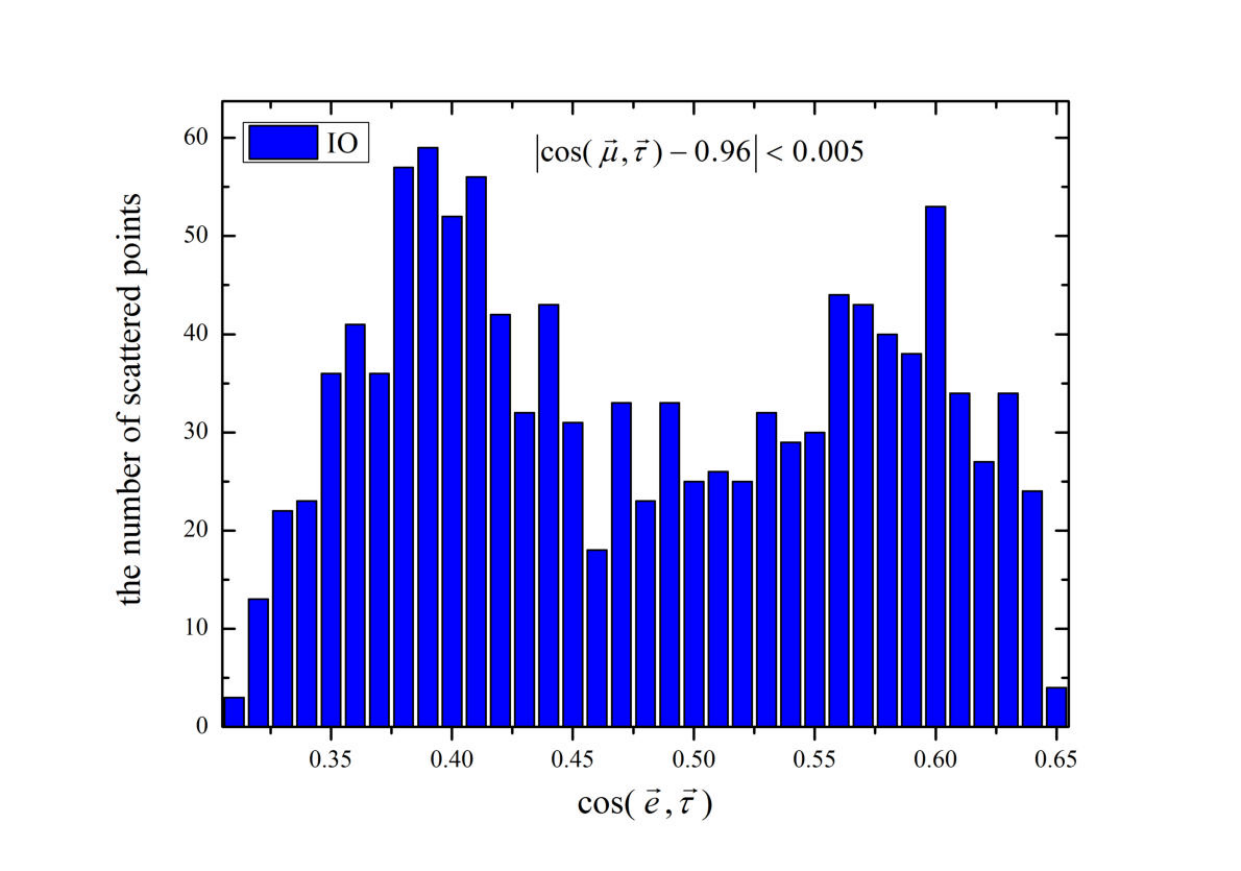}
  \caption{\label{fig:4} The number of scattered points corresponding to a series of small intervals of $\cos(\protect\overrightarrow{e},\protect\overrightarrow{\mu})$ and $\cos(\protect\overrightarrow{e},\protect\overrightarrow{\tau})$ in Fig.\ref{fig:2}, with the constraint of $|\cos(\protect\overrightarrow{\mu},\protect\overrightarrow{\tau})-0.96|<0.005$. Red: NO case, Blue: IO case.}
\end{figure}

We divide the data samples of $\cos(\overrightarrow{\mu},\overrightarrow{\tau})$ originated from  Fig.\ref{fig:2} into a series of small intervals such as
[0.825, 0.835], $\cdot\cdot\cdot$, [0.955, 0.965], $\cdot\cdot\cdot$, [0.995, 1.00],
which are shown in Fig.\ref{fig:3}.
\begin{figure}
\label{fig:3}
  \centering
  \includegraphics[width=.49\textwidth]{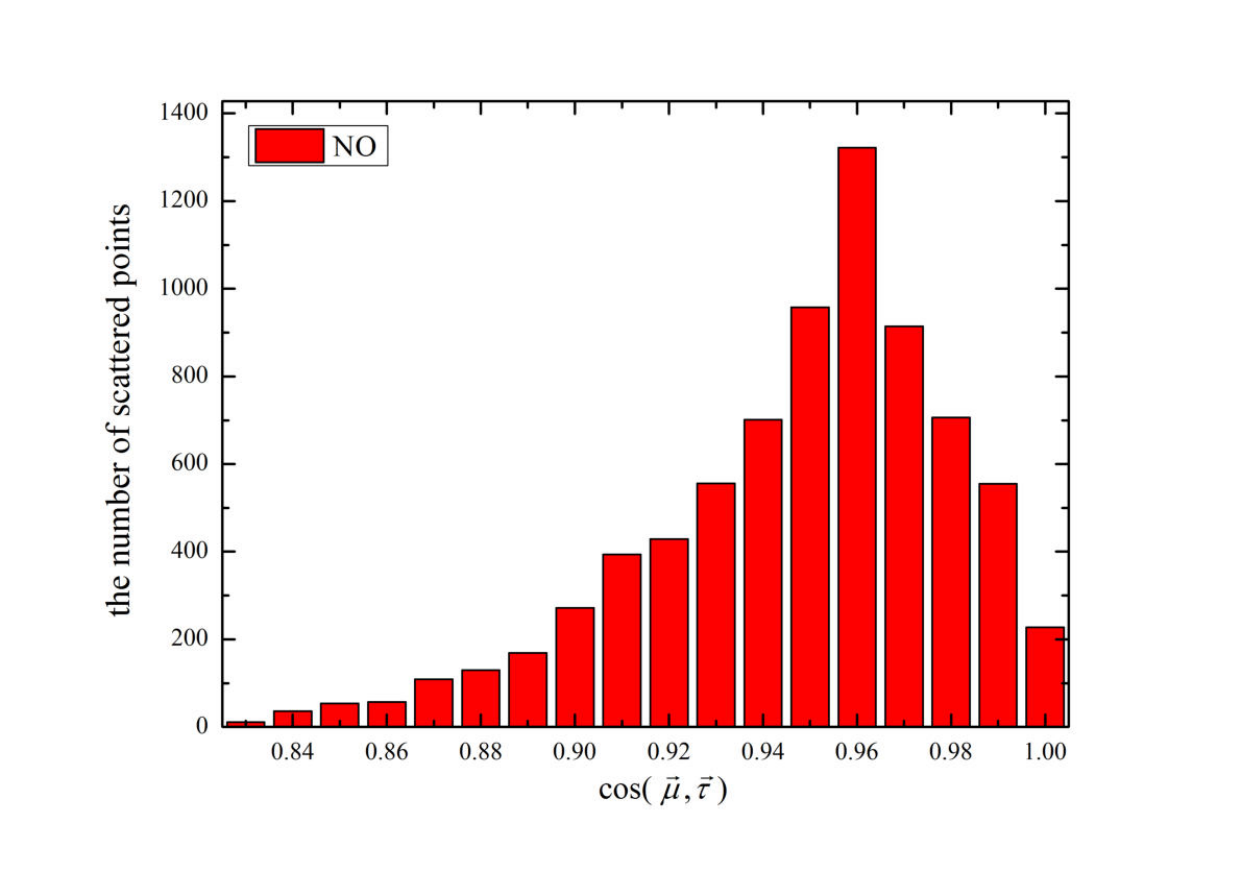}
  \hfill
  \includegraphics[width=.49\textwidth]{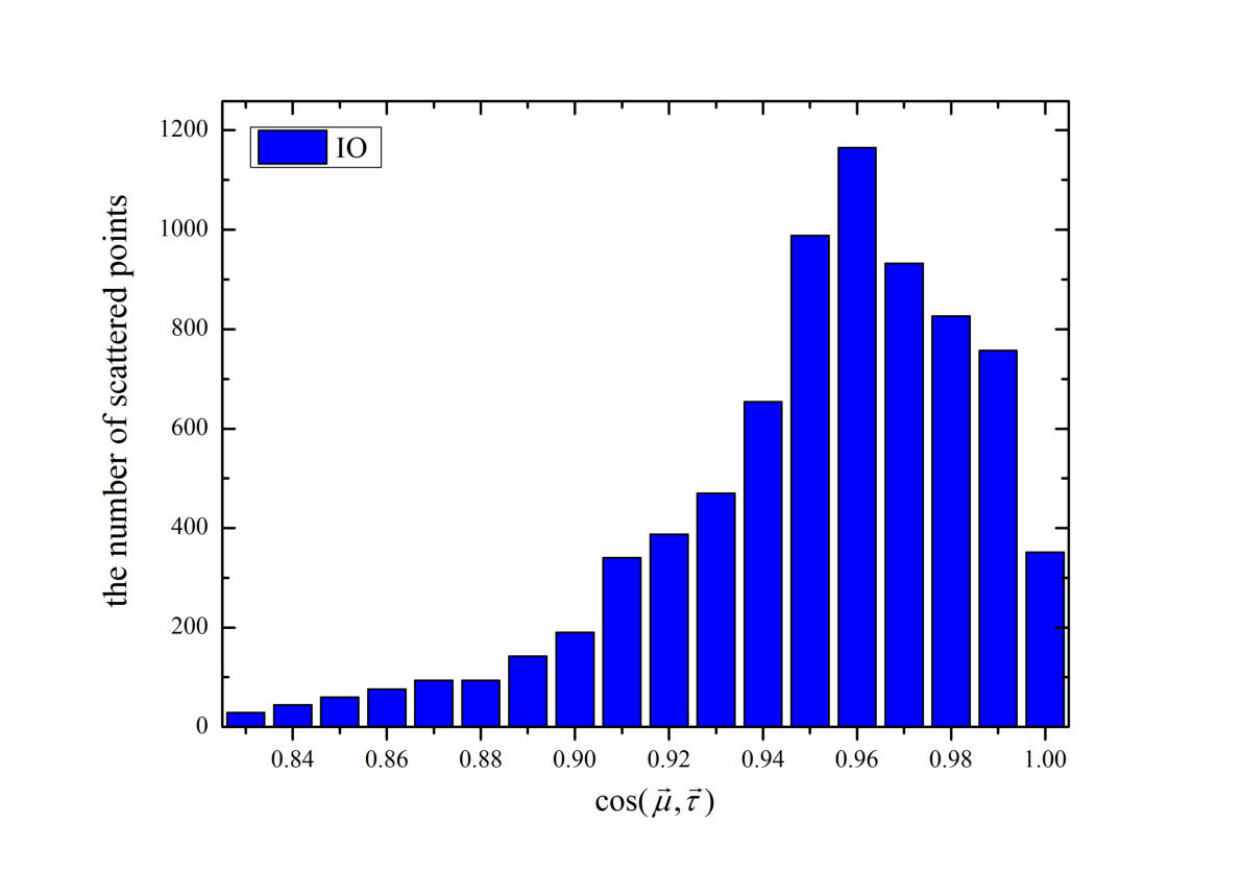}
  \caption{\label{fig:3} The number of scattered points corresponding to a series of small intervals of $\cos(\protect\overrightarrow{\mu},\protect\overrightarrow{\tau})$ in Fig.\ref{fig:2}. Red: NO case, Blue: IO case.}
\end{figure}
From Fig.\ref{fig:3}, we can observe that the number of data points with $\cos(\overrightarrow{\mu},\overrightarrow{\tau}) \in$ [0.955, 0.965] is maximal both in the NO case and IO case.
In consequence, we focus on the interval of $\cos(\overrightarrow{\mu},\overrightarrow{\tau})$, [0.955, 0.965].
Based on the similar approach, the number of data points of $\cos(\overrightarrow{e},\overrightarrow{\mu})$ and $\cos(\overrightarrow{e},\overrightarrow{\tau})$ is presented in Fig.\ref{fig:4}, with the constraint of $|\cos(\overrightarrow{\mu},\overrightarrow{\tau})-0.96|<0.005$.
From Fig.\ref{fig:4}, we can see that the large numbers of scattered points in several intervals are approximate in the IO case, e.g. the $\cos(\overrightarrow{e},\overrightarrow{\mu})$ intervals [0.575, 0.585], [0.595, 0.605], the $\cos(\overrightarrow{e},\overrightarrow{\tau})$ intervals [0.375, 0.385], [0.385, 0.395] and [0.405, 0.415].
However, the numerical results show that these intervals with the similar number of scattered points in the same dense region bring no obvious differences  on the correlations of the lepton mixing parameters.
Therefore, for the sake of illustration, we select the intervals of $\cos(\overrightarrow{e},\overrightarrow{\mu})$ and $\cos(\overrightarrow{e},\overrightarrow{\tau})$ with the largest number of scattered points in every dense area of Fig.\ref{fig:2}.

\begin{figure}
  \centering
  \includegraphics[width=.49\textwidth]{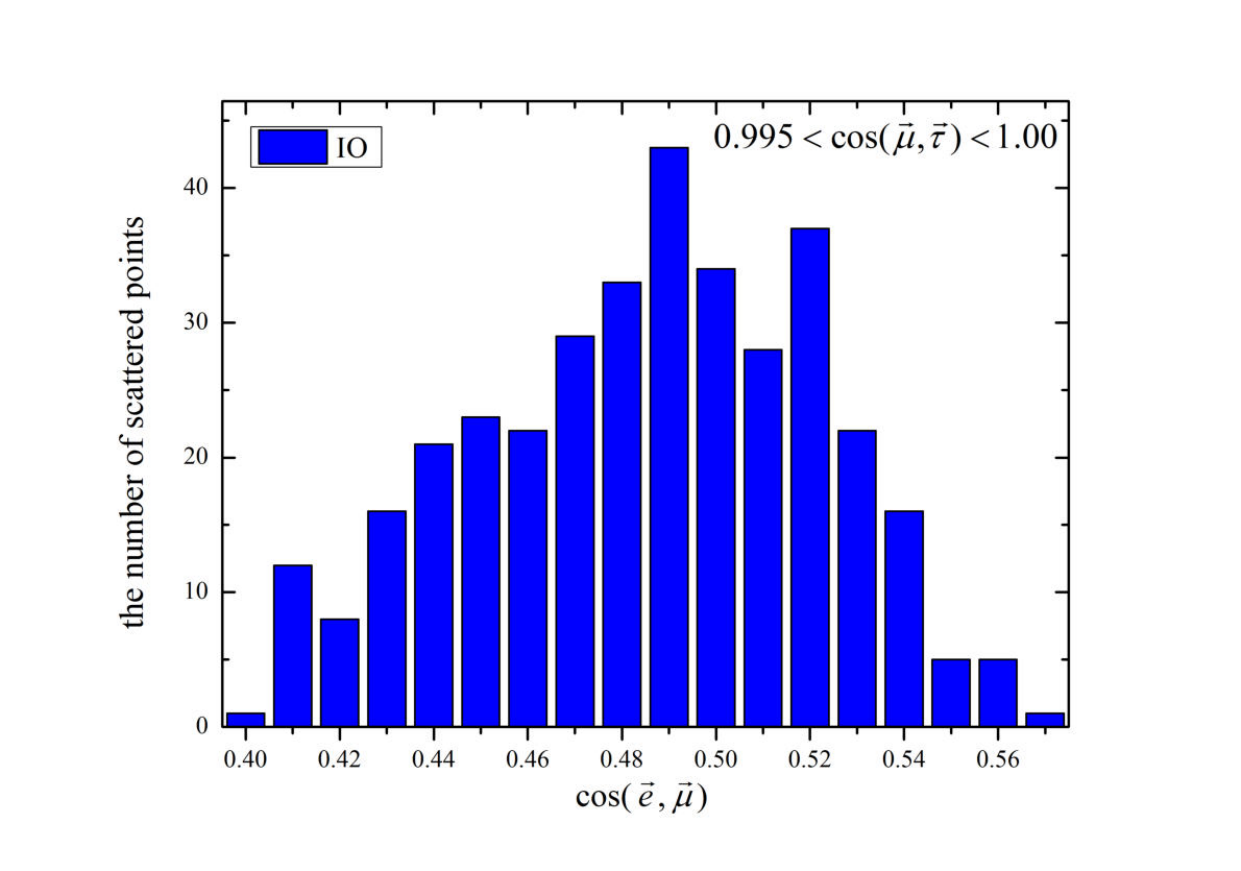}
  \hfill
  \includegraphics[width=.49\textwidth]{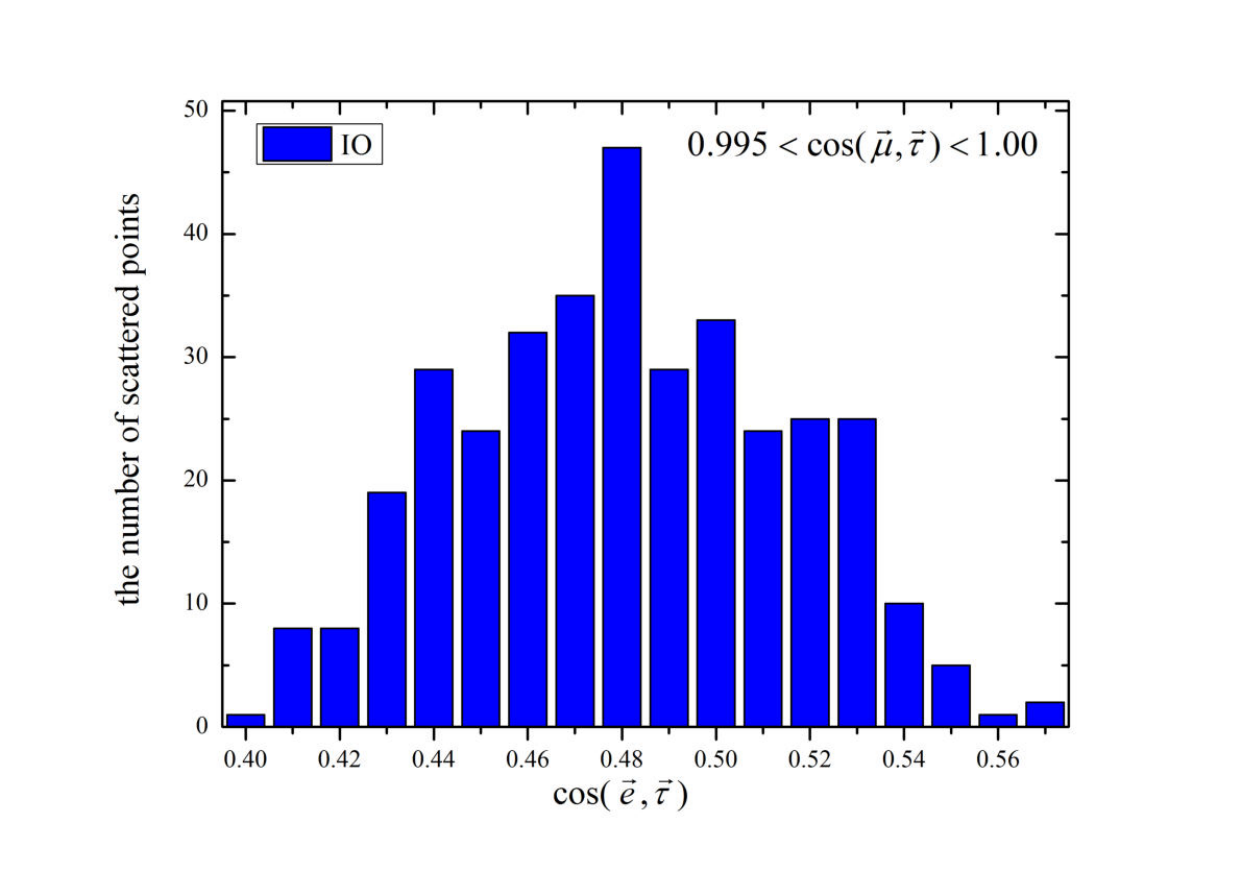}
  \caption{\label{fig:5} The number of scattered points corresponding to a series of small intervals of $\cos(\protect\overrightarrow{e},\protect\overrightarrow{\mu})$ and $\cos(\protect\overrightarrow{e},\protect\overrightarrow{\tau})$ in the IO case of Fig.\ref{fig:2}, with the constraint of $0.995<\cos(\protect\overrightarrow{\mu},\protect\overrightarrow{\tau})<1.00$.}
\end{figure}

In addition,  the approximated $\mu-\tau$ reflection symmetry corresponding to one of the dense regions in the IO case should be considered.
In this case, the number of scattered points of $\cos(\overrightarrow{e},\overrightarrow{\mu})$ and $\cos(\overrightarrow{e},\overrightarrow{\tau})$  is shown in Fig.\ref{fig:5}, with the constraint of $0.995<\cos(\overrightarrow{\mu},\overrightarrow{\tau})<1.00$.
From Fig.\ref{fig:5}, we find that the largest numbers of scattered points corresponds to the intervals $|\cos(\overrightarrow{e},\overrightarrow{\mu})-0.49|<0.005$ and $|\cos(\overrightarrow{e},\overrightarrow{\tau})-0.48|<0.005$.

On the basis of the above analysis, the selected  representative intervals of the angles between the row vectors are listed in Tab.\ref{tab:2}.
\begin{table}
\label{tab:2}
\caption{\label{tab:2} The intervals of the angles between the row vectors with the largest number of scattered points stem from every dense regions of Fig.\ref{fig:2}.}
  \centering
  \begin{tabular}{c c c c}
     \noalign{\smallskip}\hline
     \noalign{\smallskip}\hline
     ~~Mass ordering ~~&~~ $\cos(\overrightarrow{\mu},\overrightarrow{\tau})$ ~~&~~$\cos(\overrightarrow{e},\overrightarrow{\mu})$~~&~~$\cos(\overrightarrow{e},\overrightarrow{\tau})$~~ \\
     \noalign{\smallskip}\hline
     ~~\multirow{2}*{NO} ~~~&~~~[0.955, 0.965]~~~&~~~[0.395, 0.405]~~~&~~~[0.385, 0.395]~~~ \\
                         ~~~&~~~[0.955, 0.965]~~~&~~~[0.585, 0.595]~~~&~~~[0.575, 0.585]~~~ \\
     \hline
     ~~\multirow{3}*{IO} ~~~&~~~[0.955, 0.965]~~~&~~~[0.405, 0.415]~~~&~~~[0.385, 0.395]~~~ \\
                         ~~~&~~~[0.955, 0.965]~~~&~~~[0.595, 0.605]~~~&~~~[0.595, 0.605]~~~ \\
                         ~~~&~~~[0.995, 1.00]~~~&~~~[0.485, 0.495]~~~&~~~[0.475, 0.485]~~~ \\
     \hline
   \end{tabular}
\end{table}

\begin{figure}
  \centering
  \includegraphics[width=.42\textwidth,height=0.21\textheight]{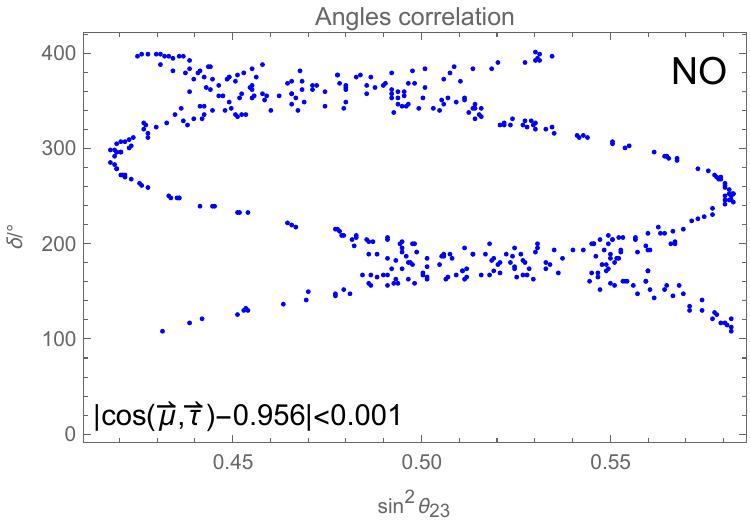}
  \hfill
  \includegraphics[width=.42\textwidth,height=0.21\textheight]{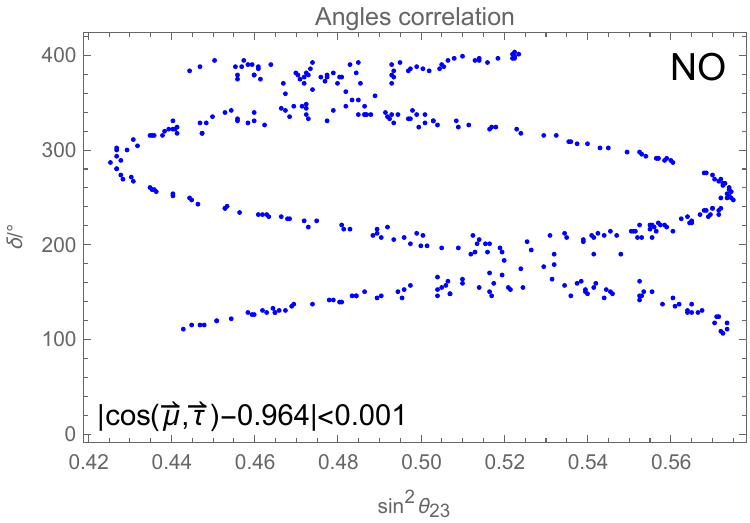}
  \hfill
  \includegraphics[width=.42\textwidth,height=0.21\textheight]{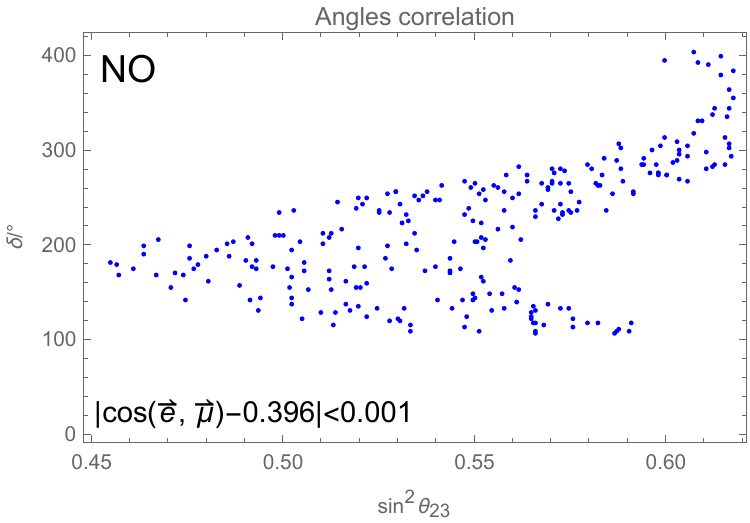}
  \hfill
  \includegraphics[width=.42\textwidth,height=0.21\textheight]{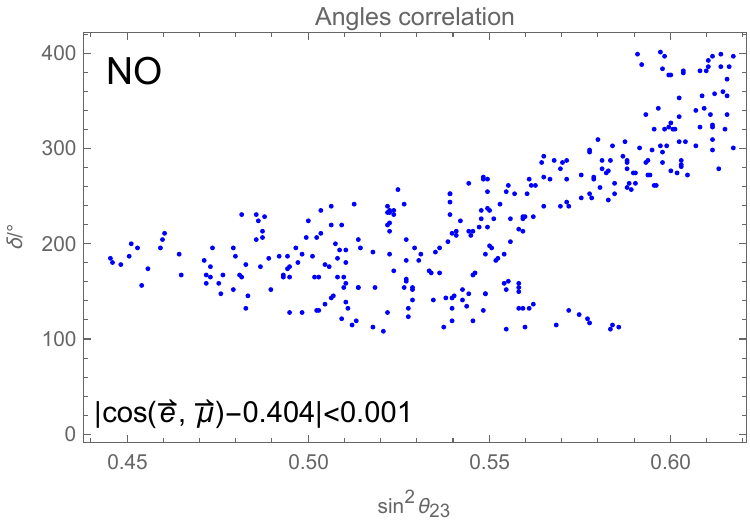}
  \hfill
  \includegraphics[width=.42\textwidth,height=0.21\textheight]{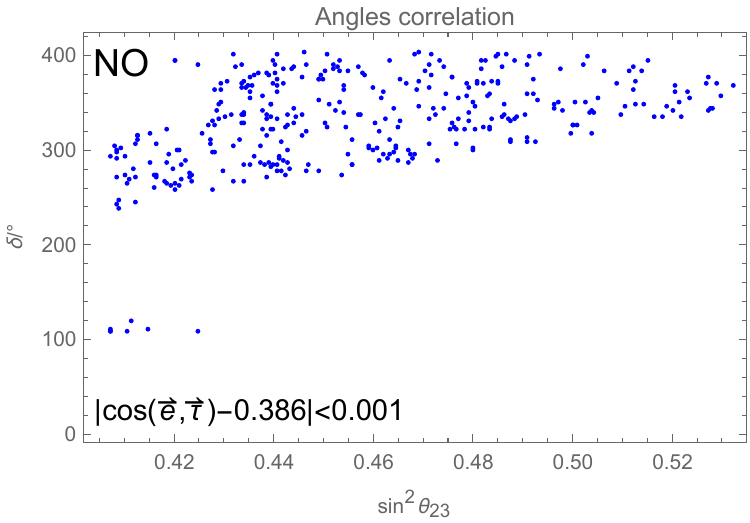}
  \hfill
  \includegraphics[width=.42\textwidth,height=0.21\textheight]{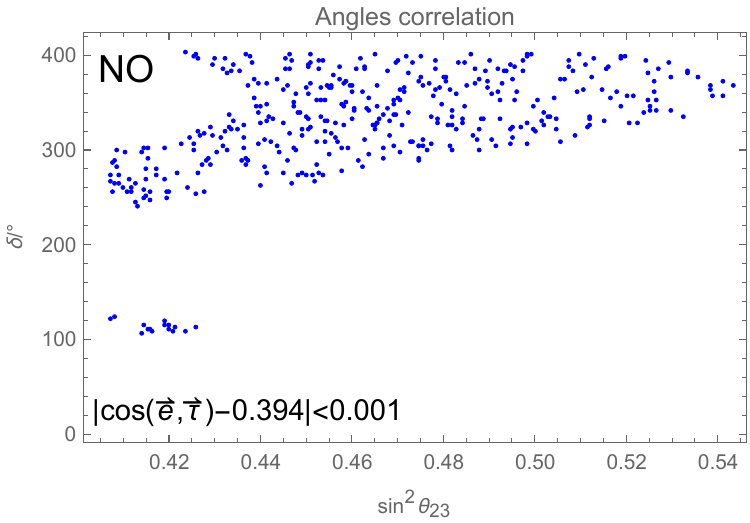}
  \caption{\label{fig:6} The correlations of $\sin^2\theta_{23}$ - $\delta$ constrained by the correlations of Eq.\ref{eq:5} - Eq.\ref{eq:7} at the $3\sigma$ level of the global fit data\cite{43}.}
\end{figure}

Now we show that the difference on the correlations of the lepton mixing parameters  from two arbitrary data points in the same interval of Fig.\ref{fig:3} - Fig.\ref{fig:5} can be ignored.
For specific explanation, three intervals in the NO case listed in Tab.\ref{tab:2} are taken, which are listed as follows: $\cos(\overrightarrow{\mu},\overrightarrow{\tau})\in$ [0.955, 0.965],
$\cos(\overrightarrow{e},\overrightarrow{\mu})\in$ [0.395, 0.405],
$\cos(\overrightarrow{e},\overrightarrow{\tau})\in$ [0.385, 0.395].
We set constraints on the leptonic mixing parameters with small neighborhoods around the data points, which are expressed as:
\begin{equation}
\label{eq:5}
|\cos(\overrightarrow{\mu},\overrightarrow{\tau})-f_{1}|<0.001,~~~~~
|\cos(\overrightarrow{\mu},\overrightarrow{\tau})-f_{2}|<0.001,
\end{equation}
\begin{equation}
\label{eq:6}
|\cos(\overrightarrow{e},\overrightarrow{\mu})-g_{1}|<0.001,~~~~~
|\cos(\overrightarrow{e},\overrightarrow{\mu})-g_{2}|<0.001,
\end{equation}
\begin{equation}
\label{eq:7}
|\cos(\overrightarrow{e},\overrightarrow{\tau})-h_{1}|<0.001,~~~~~
|\cos(\overrightarrow{e},\overrightarrow{\tau})-h_{2}|<0.001,
\end{equation}
where ($f_{1}=0.956, f_{2}=0.964$), ($g_{1}=0.396, g_{2}=0.404$), ($h_{1}=0.386, h_{2}=0.394$) are two data points near the boundary of the interval [0.955, 0.965], [0.395, 0.405] and [0.385, 0.395], respectively. The correlations of $\sin^2\theta_{23}$ - $\delta$  from the above constraints are obtained at the $3\sigma$ level of the global fit data\cite{43}, which are shown in Fig.\ref{fig:6}.
Here, the correlations of $\sin^2\theta_{13}$ - $\sin^2\theta_{12}$ from the constrains of Eq.\ref{eq:5} -Eq.\ref{eq:7} are not given
because they are insensitive to these constrains. In other words, the data points of the correlation of $\sin^2\theta_{13}$ - $\sin^2\theta_{12}$ are of a uniform distribution at the 3$\sigma$ level.
From Fig.\ref{fig:6}, we can see that the scattered points distribution of the correlation is not sensitive to the representative points in the interval.
Therefore, we choose the middle point as the representative of an interval. For the selected intervals listed in Tab.\ref{tab:2}, the typical points  are given in Tab.\ref{tab:3}.
\begin{table}
\label{tab:3}
\caption{\label{tab:3} Typical data points in the dense regions of Fig.\ref{fig:2}.}
  \centering
  \begin{tabular}{c c c c}
     \noalign{\smallskip}\hline
     \noalign{\smallskip}\hline
     ~~Mass ordering ~~&~~ $\cos(\overrightarrow{\mu_{0}},\overrightarrow{\tau_{0}})$ ~~&~~$\cos(\overrightarrow{e_{0}},\overrightarrow{\mu_{0}})$~~&~~$\cos(\overrightarrow{e_{0}},\overrightarrow{\tau_{0}})$~~ \\
     \noalign{\smallskip}\hline
     ~~NO ~~~&~~~0.96~~~&~~~0.40, ~0.59~~~&~~~0.39, ~0.58~~~ \\
     \hline
     ~~\multirow{2}*{IO} ~~~&~~~0.96~~~&~~~0.41, ~0.60~~~&~~~0.39, ~0.60~~~ \\
                         ~~~&~~~0.997~~~&~~~0.49~~~&~~~0.48~~~ \\
     \hline
   \end{tabular}
\end{table}

\subsection{Leptonic mixing parameters constrained by the geometric correlations}
Now we analyse the impacts of  the geometric correlations on the leptonic mixing patterns.
\subsubsection{Impacts of the correlations of two included angles}
\begin{figure}
  \centering
  \includegraphics[width=.42\textwidth,height=0.21\textheight]{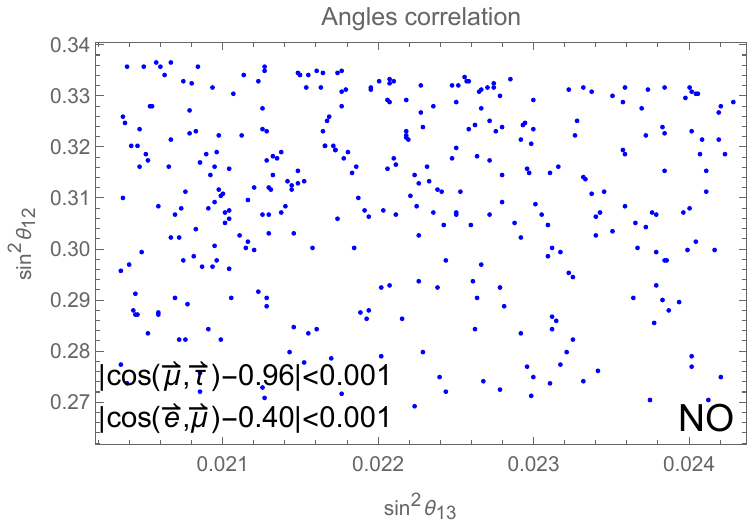}
  \hfill
  \includegraphics[width=.42\textwidth,height=0.21\textheight]{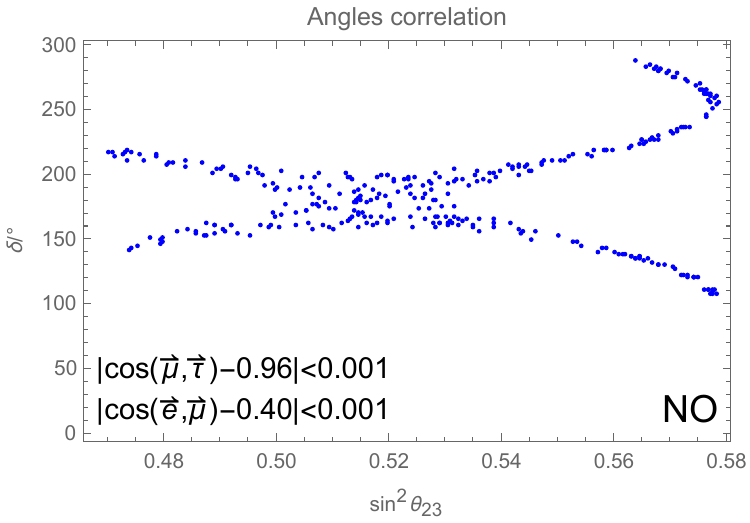}
  \hfill
  \includegraphics[width=.42\textwidth,height=0.21\textheight]{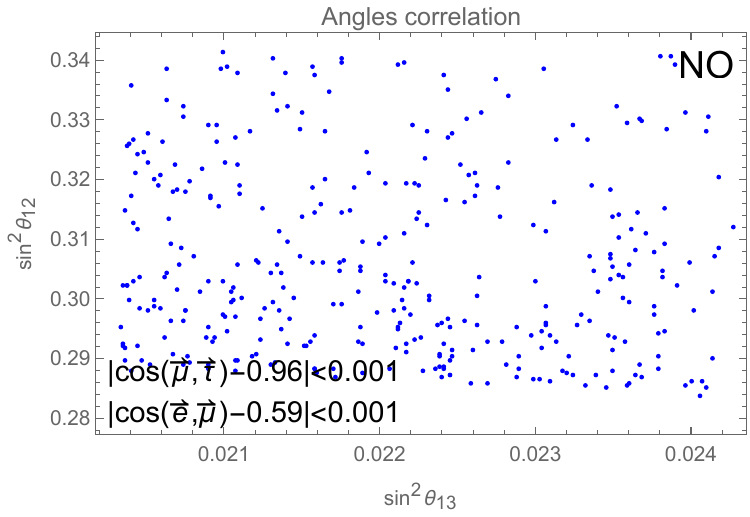}
  \hfill
  \includegraphics[width=.42\textwidth,height=0.21\textheight]{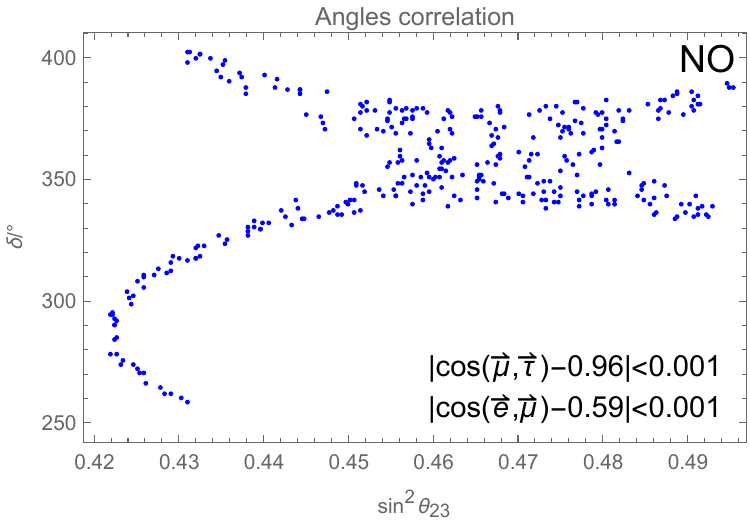}
  \hfill
  \includegraphics[width=.42\textwidth,height=0.21\textheight]{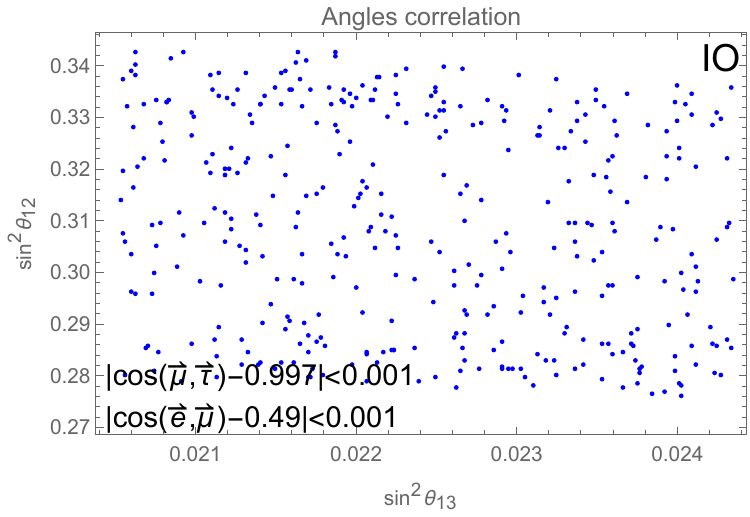}
  \hfill
  \includegraphics[width=.42\textwidth,height=0.21\textheight]{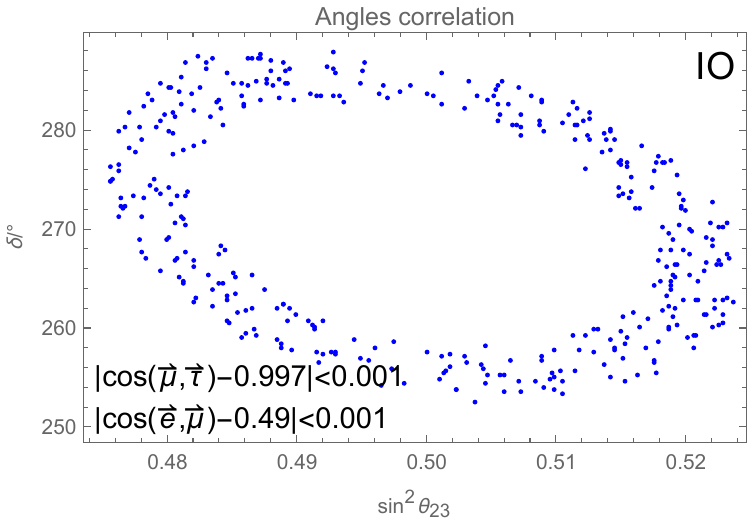}
  \caption{\label{fig:7} Leptonic mixing parameters constrained by the correlations between two included angles of (~$\cos(\protect\overrightarrow{\mu_{0}},\protect\overrightarrow{\tau_{0}}),~  \cos(\protect\overrightarrow{e_{0}},\protect\overrightarrow{\mu_{0}})$~) listed in Tab.\ref{tab:4}. Top: NO case with the constraints of $|\cos(\protect\overrightarrow{\mu},\protect\overrightarrow{\tau})-0.96|<0.001$ and $|\cos(\protect\overrightarrow{e},\protect\overrightarrow{\mu})-0.40|<0.001$. Center: NO case with the constraints of $|\cos(\protect\overrightarrow{\mu},\protect\overrightarrow{\tau})-0.96|<0.001$ and $|\cos(\protect\overrightarrow{e},\protect\overrightarrow{\mu})-0.59|<0.001$. Bottom: IO case with the constraints of $|\cos(\protect\overrightarrow{\mu},\protect\overrightarrow{\tau})-0.997|<0.001$ and $|\cos(\protect\overrightarrow{e},\protect\overrightarrow{\mu})-0.49|<0.001$.}
\end{figure}

\begin{figure}
  \centering
  \includegraphics[width=.42\textwidth,height=0.21\textheight]{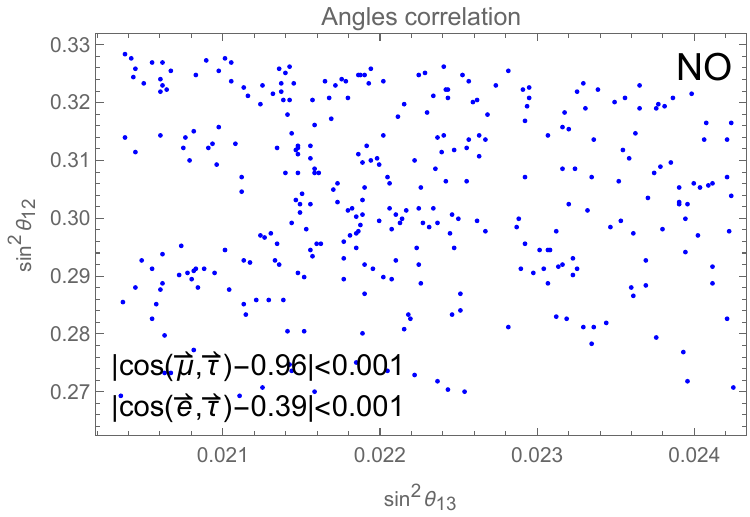}
  \hfill
  \includegraphics[width=.42\textwidth,height=0.21\textheight]{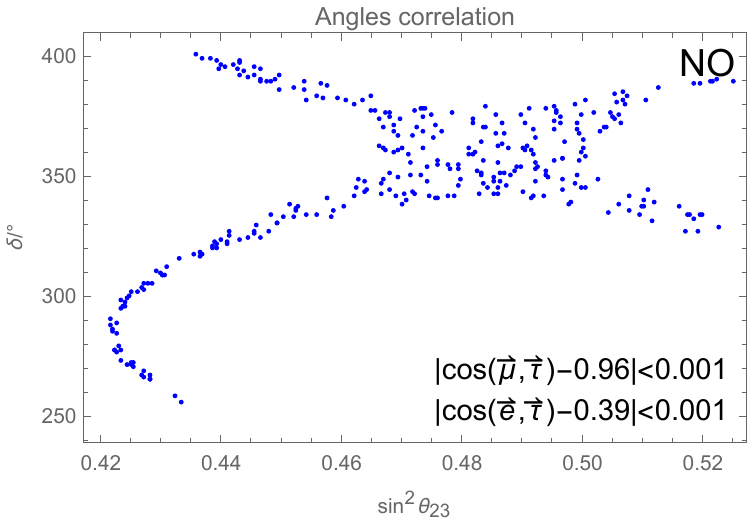}
  \hfill
  \includegraphics[width=.42\textwidth,height=0.21\textheight]{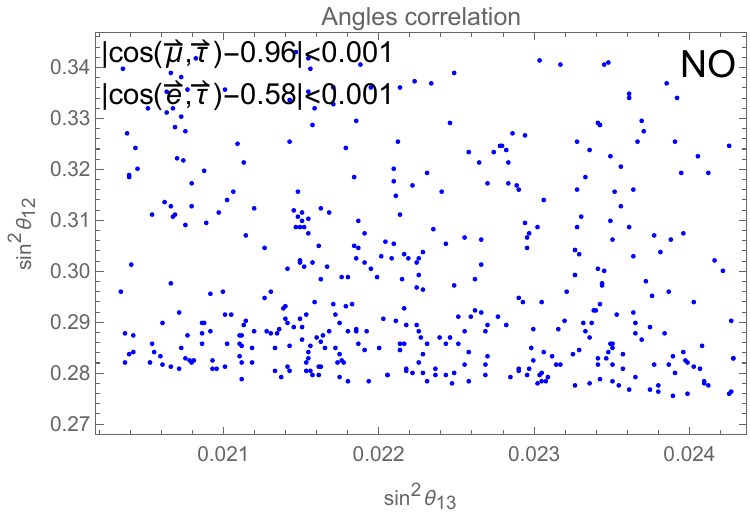}
  \hfill
  \includegraphics[width=.42\textwidth,height=0.21\textheight]{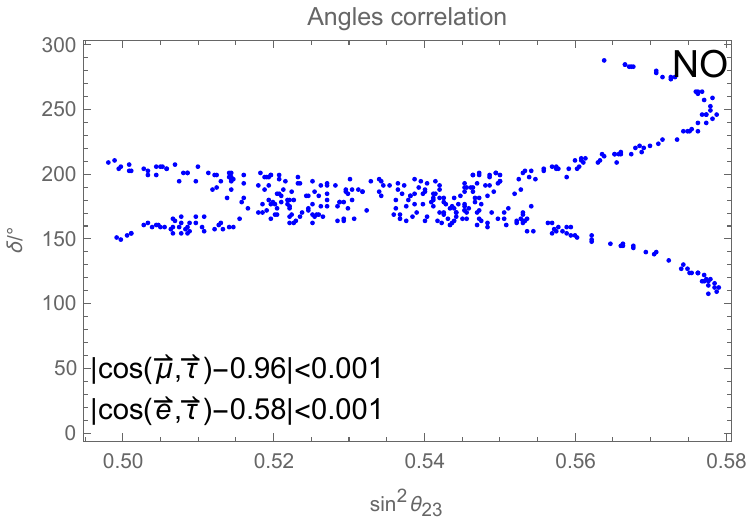}
  \hfill
  \includegraphics[width=.42\textwidth,height=0.21\textheight]{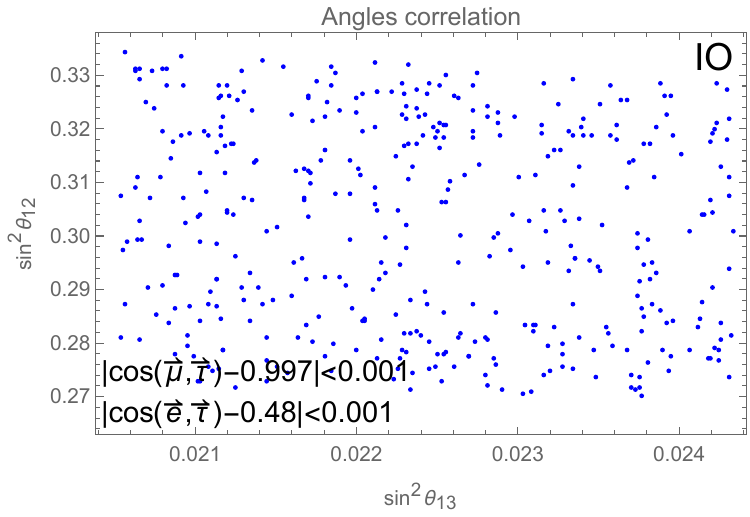}
  \hfill
  \includegraphics[width=.42\textwidth,height=0.21\textheight]{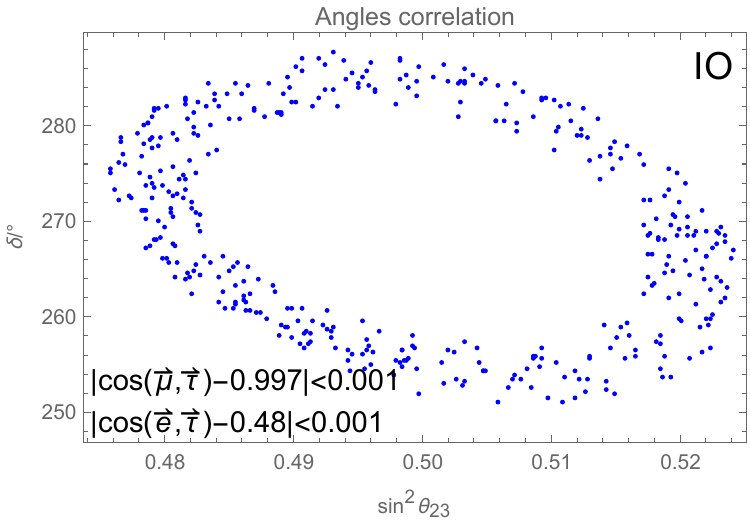}
  \caption{\label{fig:8} Leptonic mixing parameters constrained by the correlations between two included angles of (~$\cos(\protect\overrightarrow{\mu_{0}},\protect\overrightarrow{\tau_{0}}),~  \cos(\protect\overrightarrow{e_{0}},\protect\overrightarrow{\tau_{0}})$~) listed in Tab.\ref{tab:4}. Top: NO case with the constraints of $|\cos(\protect\overrightarrow{\mu},\protect\overrightarrow{\tau})-0.96|<0.001$ and $|\cos(\protect\overrightarrow{e},\protect\overrightarrow{\tau})-0.39|<0.001$. Center: NO case with the constraints of $|\cos(\protect\overrightarrow{\mu},\protect\overrightarrow{\tau})-0.96|<0.001$ and $|\cos(\protect\overrightarrow{e},\protect\overrightarrow{\tau})-0.58|<0.001$. Bottom: IO case with the constraints of $|\cos(\protect\overrightarrow{\mu},\protect\overrightarrow{\tau})-0.997|<0.001$ and $|\cos(\protect\overrightarrow{e},\protect\overrightarrow{\tau})-0.48|<0.001$.}
\end{figure}

\begin{figure}
  \centering
  \includegraphics[width=.42\textwidth,height=0.21\textheight]{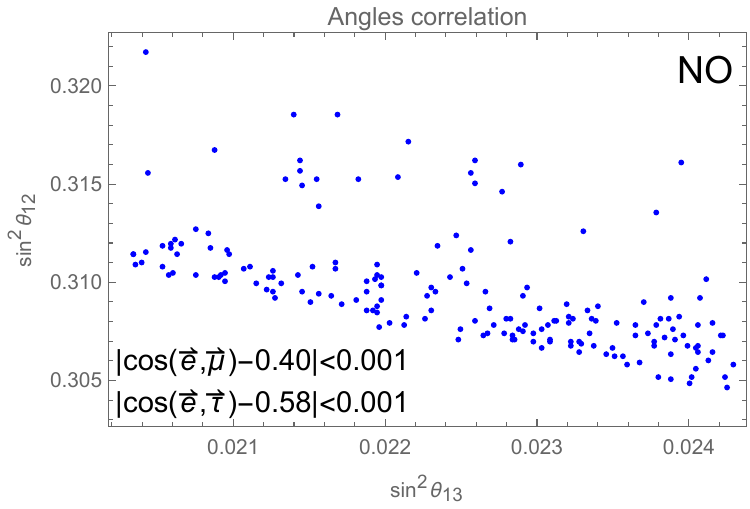}
  \hfill
  \includegraphics[width=.42\textwidth,height=0.21\textheight]{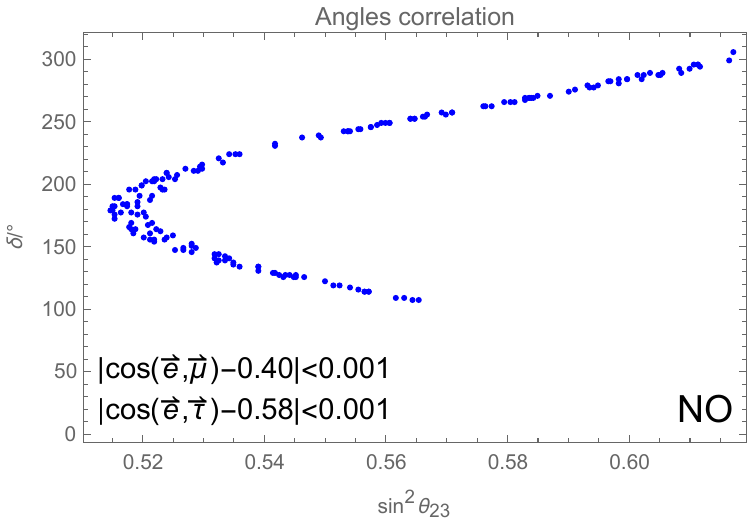}
  \hfill
  \includegraphics[width=.42\textwidth,height=0.21\textheight]{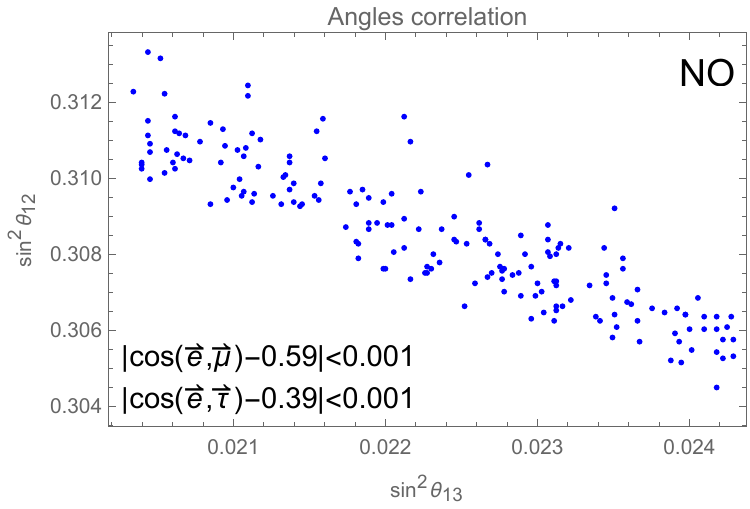}
  \hfill
  \includegraphics[width=.42\textwidth,height=0.21\textheight]{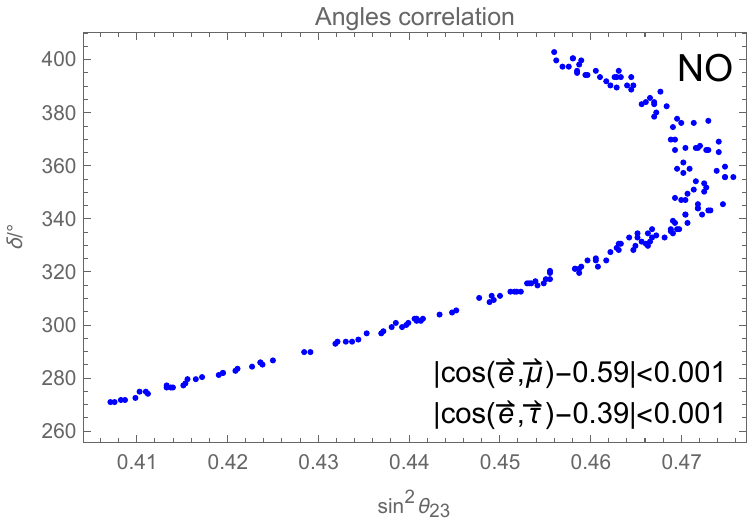}
  \hfill
  \includegraphics[width=.42\textwidth,height=0.21\textheight]{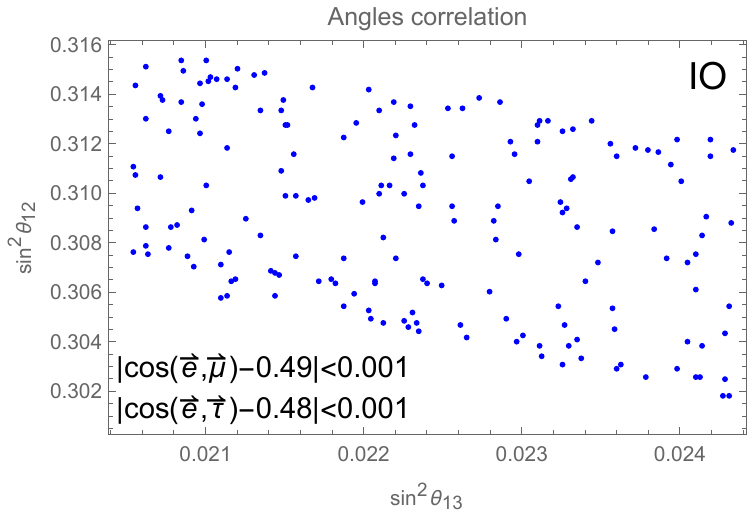}
  \hfill
  \includegraphics[width=.42\textwidth,height=0.21\textheight]{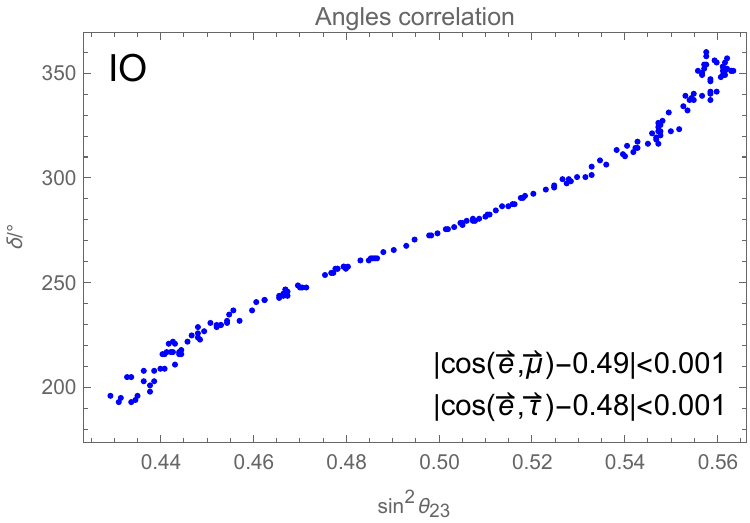}
  \caption{\label{fig:9} Leptonic mixing parameters constrained by the correlations between two included angles of (~$\cos(\protect\overrightarrow{e_{0}},\protect\overrightarrow{\mu_{0}}),~  \cos(\protect\overrightarrow{e_{0}},\protect\overrightarrow{\tau_{0}})$~) listed in Tab.\ref{tab:4}. Top: NO case with the constraints of $|\cos(\protect\overrightarrow{e},\protect\overrightarrow{\mu})-0.40|<0.001$ and $|\cos(\protect\overrightarrow{e},\protect\overrightarrow{\tau})-0.58|<0.001$. Center: NO case with the constraints of $|\cos(\protect\overrightarrow{e},\protect\overrightarrow{\mu})-0.59|<0.001$ and $|\cos(\protect\overrightarrow{e},\protect\overrightarrow{\tau})-0.39|<0.001$. Bottom: IO case with the constraints of $|\cos(\protect\overrightarrow{e},\protect\overrightarrow{\mu})-0.49|<0.001$ and $|\cos(\protect\overrightarrow{e},\protect\overrightarrow{\tau})-0.48|<0.001$.}
\end{figure}

According to the typical points shown in  Tab.\ref{tab:3}, 15 viable correlations of two angles between the row vectors are obtained at the $3\sigma$ level of the global fit data\cite{43}, which are listed in Tab.\ref{tab:4}.
\begin{table}
\label{tab:4}
\caption{\label{tab:4} Viable combinations with two included angles of the row vectors.}
  \centering
  \begin{tabular}{c c c c}
     \noalign{\smallskip}\hline
     \noalign{\smallskip}\hline
     ~~Mass ordering ~~&~~$(\cos(\overrightarrow{\mu_{0}},\overrightarrow{\tau_{0}}),~ \cos(\overrightarrow{e_{0}},\overrightarrow{\mu_{0}}))$ ~~&~~$(\cos(\overrightarrow{\mu_{0}},\overrightarrow{\tau_{0}}),~ \cos(\overrightarrow{e_{0}},\overrightarrow{\tau_{0}}))$~~&~~$(\cos(\overrightarrow{e_{0}},\overrightarrow{\mu_{0}}),~ \cos(\overrightarrow{e_{0}},\overrightarrow{\tau_{0}}))$~~ \\
     \noalign{\smallskip}\hline
     ~~\multirow{2}*{NO} ~~~&~~~(0.96, 0.40)~~~&~~~(0.96, 0.39)~~~&~~~(0.40, 0.58)~~~ \\
                         ~~~&~~~(0.96, 0.59)~~~&~~~(0.96, 0.58)~~~&~~~(0.59, 0.39)~~~ \\
     \hline
     ~~\multirow{3}*{IO} ~~~&~~~(0.96, 0.41)~~~&~~~(0.96, 0.39)~~~&~~~(0.41, 0.60)~~~ \\
                         ~~~&~~~(0.96, 0.60)~~~&~~~(0.96, 0.60)~~~&~~~(0.60, 0.39)~~~ \\
                         ~~~&~~~(0.997, 0.49)~~~&~~~(0.997, 0.48)~~~&~~~(0.49, 0.48)~~~ \\
     \hline
   \end{tabular}
\end{table}
Employing the correlations of the included angles in Tab.\ref{tab:4}, we set the geometric constraints  as follows:
\begin{equation}
\label{eq:8}
\begin{cases}
|\cos(\overrightarrow{\mu},\overrightarrow{\tau})-\cos(\overrightarrow{\mu_{0}},\overrightarrow{\tau_{0}})|<0.001,\\
|\cos(\overrightarrow{e},\overrightarrow{\mu})-\cos(\overrightarrow{e_{0}},\overrightarrow{\mu_{0}})|<0.001,
\end{cases}
\end{equation}
\begin{equation}
\label{eq:9}
\begin{cases}
|\cos(\overrightarrow{\mu},\overrightarrow{\tau})-\cos(\overrightarrow{\mu_{0}},\overrightarrow{\tau_{0}})|<0.001,\\
|\cos(\overrightarrow{e},\overrightarrow{\tau})-\cos(\overrightarrow{e_{0}},\overrightarrow{\tau_{0}})|<0.001,
\end{cases}
\end{equation}
\begin{equation}
\label{eq:10}
\begin{cases}
|\cos(\overrightarrow{e},\overrightarrow{\mu})-\cos(\overrightarrow{e_{0}},\overrightarrow{\mu_{0}})|<0.001,\\
|\cos(\overrightarrow{e},\overrightarrow{\tau})-\cos(\overrightarrow{e_{0}},\overrightarrow{\tau_{0}})|<0.001.
\end{cases}
\end{equation}
On the bases of the constraints, the correlations between the leptonic mixing parameters are shown in Fig.\ref{fig:7} - Fig.\ref{fig:9}.
Since the figure in the IO case is similar to the counterpart in the NO case, the plots of correlations in the IO case except the plots for the approximated $\mu-\tau$ symmetry are not shown here.
We make some comments on the main observations as follows.\\
(a) ~The range of $\sin^2\theta_{12}$ is obviously reduced by the constraint of (~$\cos(\overrightarrow{e_{0}},\overrightarrow{\mu_{0}})$, $\cos(\overrightarrow{e_{0}},\overrightarrow{\tau_{0}})$~), see Fig.\ref{fig:9}. In contrast, the influence of the constraints of (~$\cos(\overrightarrow{\mu_{0}},\overrightarrow{\tau_{0}})$, $\cos(\overrightarrow{e_{0}},\overrightarrow{\mu_{0}})$~) and (~$\cos(\overrightarrow{\mu_{0}},\overrightarrow{\tau_{0}})$, $\cos(\overrightarrow{e_{0}},\overrightarrow{\tau_{0}})$~) on the range of $\sin^2\theta_{12}$ can be neglected.
\\
(b) ~$\delta$ and $\sin^2\theta_{23}$ are sensitive to the constraint of  (~$\cos(\overrightarrow{\mu_{0}},\overrightarrow{\tau_{0}})$, $\cos(\overrightarrow{e_{0}},\overrightarrow{\mu_{0}})$~) and/or (~$\cos(\overrightarrow{\mu_{0}},\overrightarrow{\tau_{0}})$, $\cos(\overrightarrow{e_{0}},\overrightarrow{\tau_{0}})$~), that is, the ranges of the parameters are compressed apparently.\\
(c) ~The above observations are stable under the variation of the intensity of the constraint, such as
$|\cos(\overrightarrow{\alpha},\overrightarrow{\beta})-\cos(\overrightarrow{\alpha_{0}},\overrightarrow{\beta_{0}})|<0.001
\longrightarrow|\cos(\overrightarrow{\alpha},\overrightarrow{\beta})-\cos(\overrightarrow{\alpha_{0}},\overrightarrow{\beta_{0}})|<0.02$ with $\alpha, \beta=e, \mu, \tau$.
 \\
(d) ~The correlations (~$\cos(\overrightarrow{\mu_{0}},\overrightarrow{\tau_{0}})$, $\cos(\overrightarrow{e_{0}},\overrightarrow{\mu_{0}})$~)$ = $(0.96, 0.40) and (~$\cos(\overrightarrow{\mu_{0}},\overrightarrow{\tau_{0}})$, $\cos(\overrightarrow{e_{0}},\overrightarrow{\mu_{0}})$~)$ = $(0.96, 0.59)
can be converted to each other through the approximated $\mu-\tau$ interchange, which means that
$\theta_{23}\longrightarrow\pi/2-\theta_{23}$, $\delta\longrightarrow\delta+\pi$.
This conclusion also applies to the correlations (~$\cos(\overrightarrow{\mu_{0}},\overrightarrow{\tau_{0}})$, $\cos(\overrightarrow{e_{0}},\overrightarrow{\tau_{0}})$~)$ = $(0.96, 0.39), (~$\cos(\overrightarrow{\mu_{0}},\overrightarrow{\tau_{0}})$, $\cos(\overrightarrow{e_{0}},\overrightarrow{\tau_{0}})$~)$ = $(0.96, 0.58) and the correlations (~$\cos(\overrightarrow{e_{0}},\overrightarrow{\mu_{0}})$, $\cos(\overrightarrow{e_{0}},\overrightarrow{\tau_{0}})$~)$ = $(0.40, 0.58), (~$\cos(\overrightarrow{e_{0}},\overrightarrow{\mu_{0}})$, $\cos(\overrightarrow{e_{0}},\overrightarrow{\tau_{0}})$~)$ = $(0.59, 0.39).\\
(e) ~The correlation of the leptonic mixing parameters obtained from the correlation (~$\cos(\overrightarrow{\mu_{0}},\overrightarrow{\tau_{0}})$, $\cos(\overrightarrow{e_{0}},\overrightarrow{\mu_{0}})$~)$ = $(0.96, 0.40) is similar as that obtained from the  correlation (~$\cos(\overrightarrow{\mu_{0}},\overrightarrow{\tau_{0}})$, $\cos(\overrightarrow{e_{0}},\overrightarrow{\tau_{0}})$~)$ = $(0.96, 0.58).
The difference of these two correlations is that the range of  $\sin^2\theta_{23}$ from the latter is moderately compressed.
This observation also applies to the correlations (~$\cos(\overrightarrow{\mu_{0}},\overrightarrow{\tau_{0}})$, $\cos(\overrightarrow{e_{0}},\overrightarrow{\mu_{0}})$~)$ = $(0.96, 0.59) and (~$\cos(\overrightarrow{\mu_{0}},\overrightarrow{\tau_{0}})$, $\cos(\overrightarrow{e_{0}},\overrightarrow{\tau_{0}})$~)$ = $(0.96, 0.39).

\subsubsection{Leptonic mixing parameters constrained by correlations of three included angles}
According to the observations in the case of the correlations of two included angles, we find that the mixing parameters ($\sin^2\theta_{12}$,  $\sin^2\theta_{23}$, $\delta$)
are strictly constrained by the geometric correlations. Now we extract the correlations between the mixing parameters from the correlations of three included angles.

\begin{figure}
  \centering
  \includegraphics[width=.40\textwidth,height=0.19\textheight]{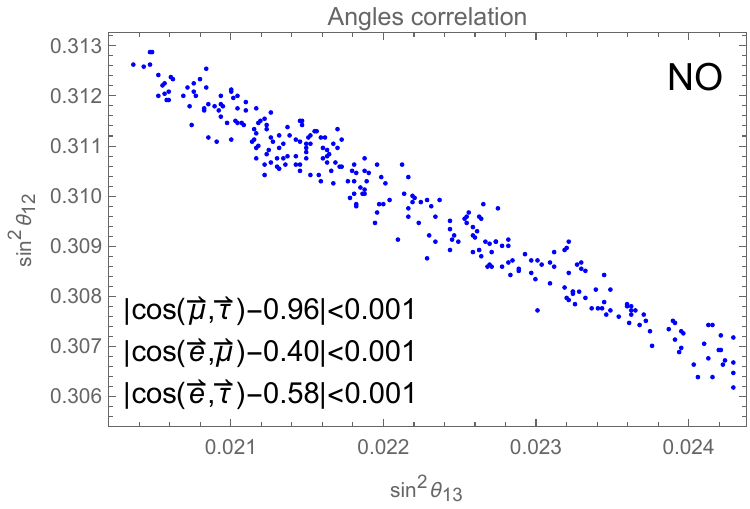}
  \hfill
  \includegraphics[width=.40\textwidth,height=0.19\textheight]{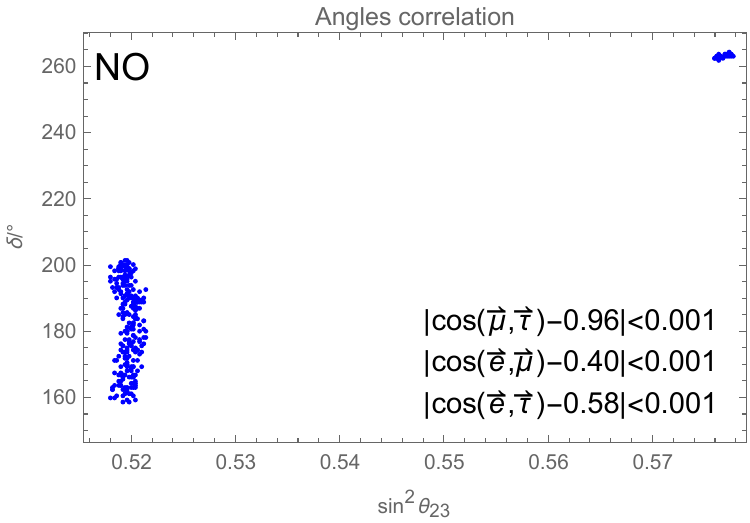}
  \hfill
  \includegraphics[width=.40\textwidth,height=0.19\textheight]{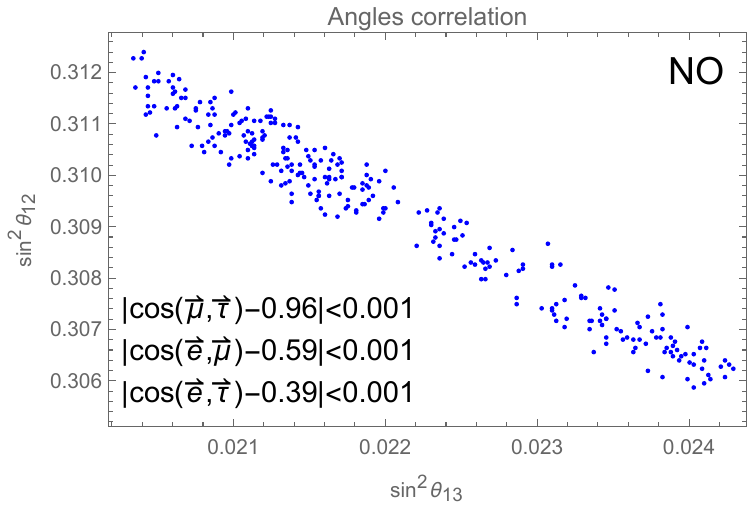}
  \hfill
  \includegraphics[width=.40\textwidth,height=0.19\textheight]{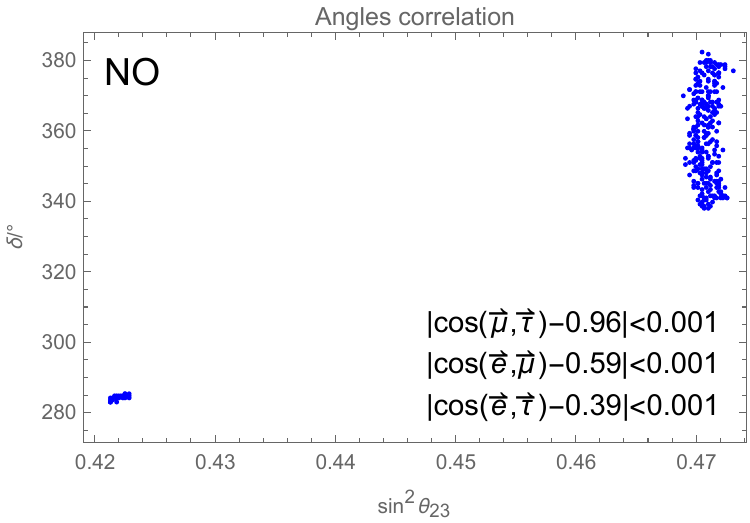}
  \hfill
  \includegraphics[width=.40\textwidth,height=0.19\textheight]{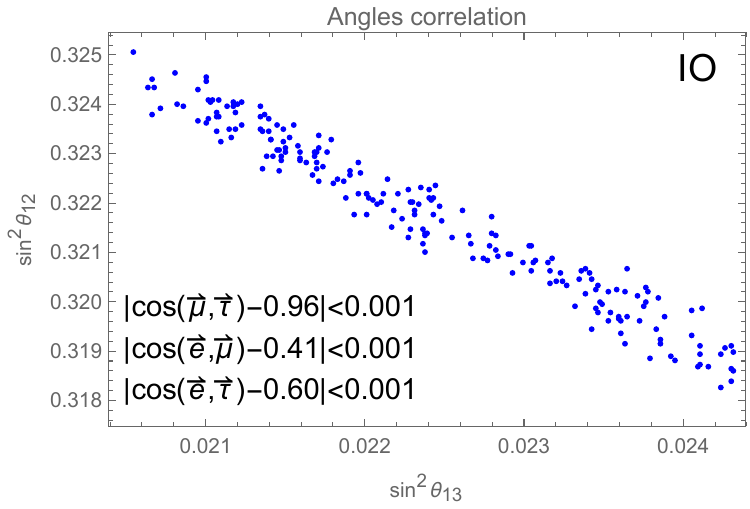}
  \hfill
  \includegraphics[width=.40\textwidth,height=0.19\textheight]{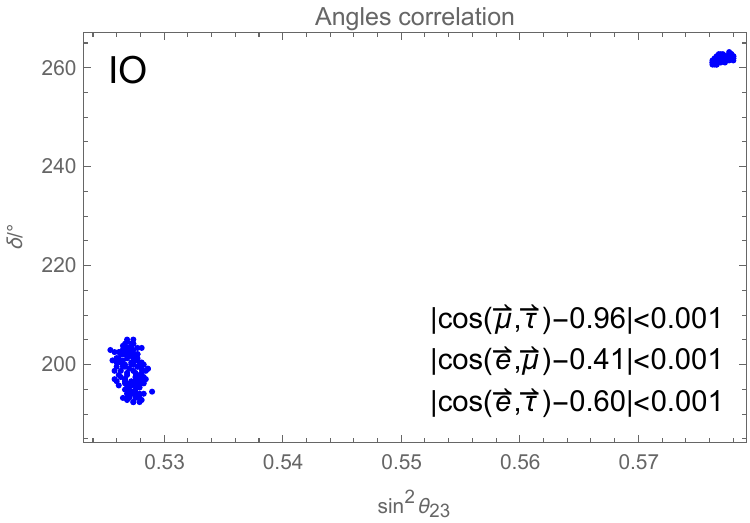}
  \hfill
  \includegraphics[width=.40\textwidth,height=0.19\textheight]{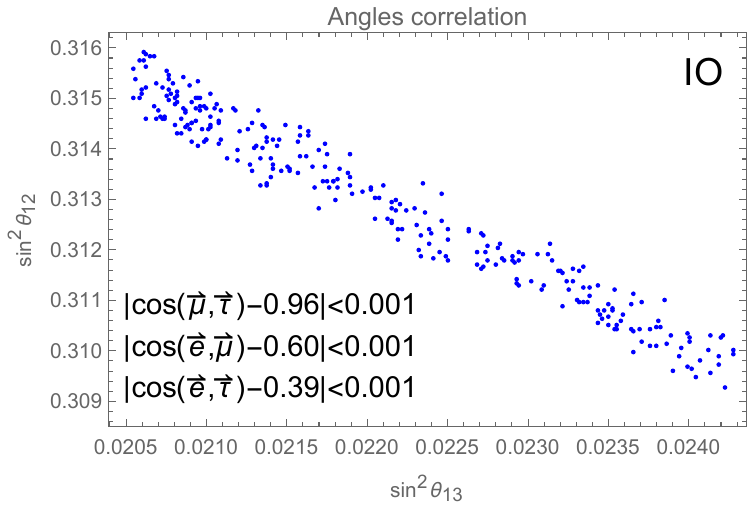}
  \hfill
  \includegraphics[width=.40\textwidth,height=0.19\textheight]{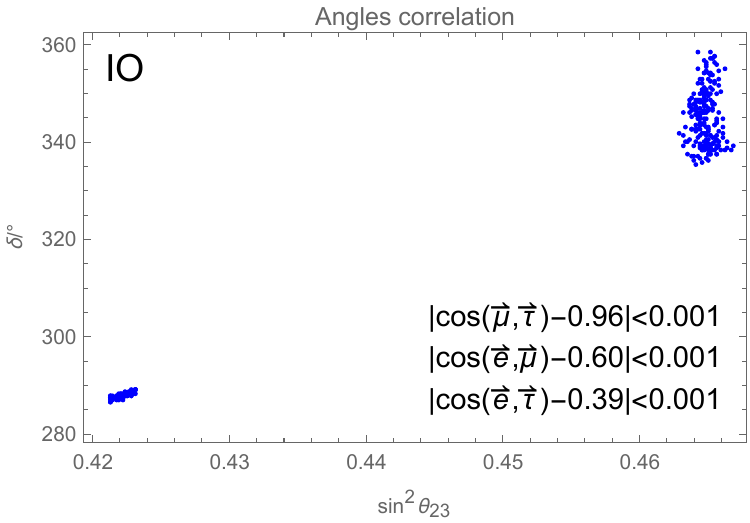}
  \hfill
  \includegraphics[width=.40\textwidth,height=0.19\textheight]{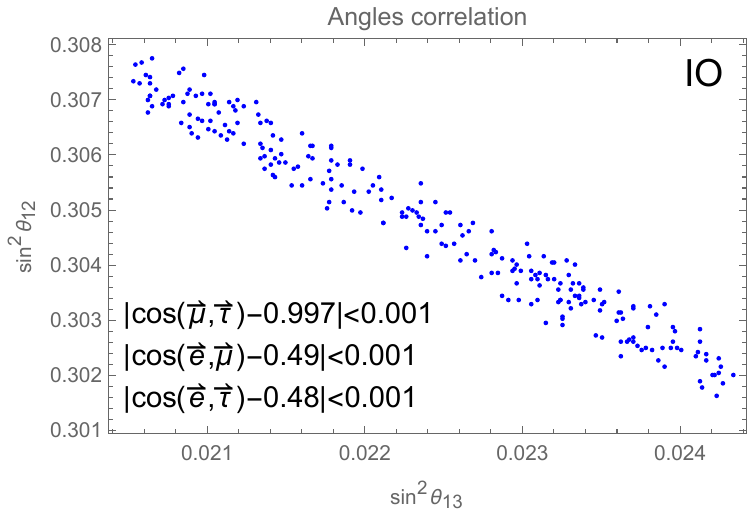}
  \hfill
  \includegraphics[width=.40\textwidth,height=0.19\textheight]{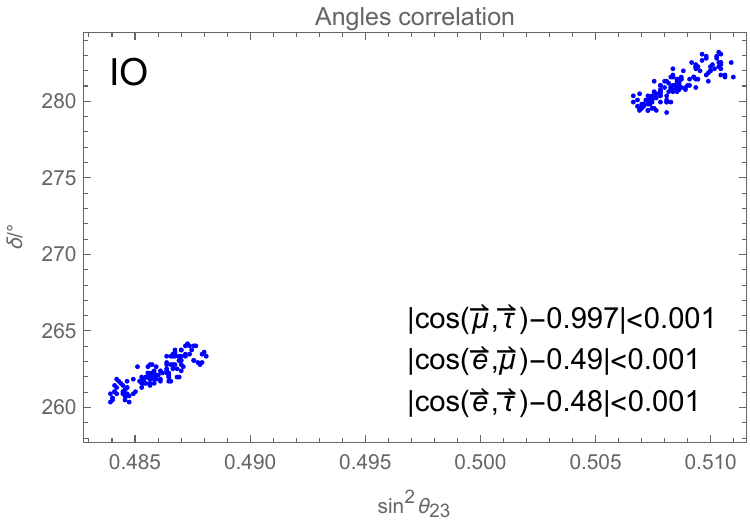}
  \caption{\label{fig:10} The correlations of $\sin^{2}\theta_{13}$ - $\sin^{2}\theta_{12}$ and $\sin^{2}\theta_{23}$ - $\delta$ from the constraints of the geometric correlations \uppercase\expandafter{\romannumeral1} - \uppercase\expandafter{\romannumeral5}.}
\end{figure}

On the basis of  Tab.\ref{tab:3}, 5 correlations of three included angles are viable at the $3\sigma$ level of the global fit data\cite{43}, which are listed in Tab.\ref{tab:5}.
\begin{table}
  \centering
  \caption{\label{tab:5} Representative correlations of three included angles.}
  \begin{tabular}{c c c}
  \noalign{\smallskip}\hline
  \noalign{\smallskip}\hline
     ~~Mass ordering ~~~&~~~Correlations~~~~&~~~~(~$\cos(\overrightarrow{\mu_{0}},\overrightarrow{\tau_{0}}),   \cos(\overrightarrow{e_{0}},\overrightarrow{\mu_{0}}),  \cos(\overrightarrow{e_{0}},\overrightarrow{\tau_{0}})$~)~~\\
     \noalign{\smallskip}\hline
     ~~\multirow{2}*{NO}~~~&~~~\uppercase\expandafter{\romannumeral1}~~~~&~~~~(0.96, 0.40, 0.58)~~\\
                        ~~~&~~~\uppercase\expandafter{\romannumeral2}~~~~&~~~~(0.96, 0.59, 0.39)~~\\
     \hline
     ~~\multirow{3}*{IO}~~~&~~~\uppercase\expandafter{\romannumeral3}~~~~&~~~~(0.96, 0.41, 0.60)~~\\
                        ~~~&~~~\uppercase\expandafter{\romannumeral4}~~~~&~~~~(0.96, 0.60, 0.39)~~\\
                        ~~~&~~~\uppercase\expandafter{\romannumeral5}~~~~&~~~~(0.997, 0.49, 0.48)~~\\
     \hline
   \end{tabular}
\end{table}
Employing the geometric constraints listed as follows:
\begin{equation}
\label{eq:11}
\begin{cases}
|\cos(\overrightarrow{\mu},\overrightarrow{\tau})-\cos(\overrightarrow{\mu_{0}},\overrightarrow{\tau_{0}})|<0.001,\\
|\cos(\overrightarrow{e},\overrightarrow{\mu})-\cos(\overrightarrow{e_{0}},\overrightarrow{\mu_{0}})|<0.001,\\
|\cos(\overrightarrow{e},\overrightarrow{\tau})-\cos(\overrightarrow{e_{0}},\overrightarrow{\tau_{0}})|<0.001,
\end{cases}
\end{equation}
we get the correlations of the leptonic mixing parameters, shown in Fig.\ref{fig:10}.
From Fig.\ref{fig:10}, we obtain the following observations:\\
(a) ~$\sin^2\theta_{12}$ is linearly dependent on $\sin^2\theta_{13}$ for all the correlations and is consistent with the recent global best fit value\cite{43} under the correlation \uppercase\expandafter{\romannumeral5}. Correspondingly, the range of $\sin^2\theta_{12}$ is reduced significantly by the geometric constraints.
Furthermore, under the constraint intensity, namely 10$^{-3}$, the variation of $\sin^2\theta_{12}$ is at most $2\times10^{-3}$. Therefore, the breaking of the $\mu-\tau$ reflection symmetry is mainly determined
by the included angles, i.e., $\cos(\overrightarrow{\alpha_{0}},\overrightarrow{\beta_{0}})$ with $\alpha$, $\beta$=$e, \mu, \tau$. \\
(b) ~The range of $\sin^2\theta_{23}$ displays two branches for all the correlations.
The right branch of $\sin^2\theta_{23}$ from the correlation \uppercase\expandafter{\romannumeral3} is consistent with the recent global best fit value\cite{43}, while the left branch of $\sin^2\theta_{23}$ from correlation \uppercase\expandafter{\romannumeral3} is consistent with the best fit value of T2K experiment ($\sin^2\theta_{23} = 0.53^{+0.03}_{-0.04}$)\cite{44}.\\
(c) ~The range of $\delta$ also presents two branches for the correlations \uppercase\expandafter{\romannumeral1}-\uppercase\expandafter{\romannumeral5}.
The best global fit value of $\delta$ in the NO case \cite{43} can be realized from the correlation \uppercase\expandafter{\romannumeral1} and the value in the IO case  can be obtained from the correlation \uppercase\expandafter{\romannumeral4} and \uppercase\expandafter{\romannumeral5}.\\
(d) ~The correlations \uppercase\expandafter{\romannumeral1} and \uppercase\expandafter{\romannumeral2} can be converted to each other through the approximated $\mu-\tau$ interchange.\\
(e) ~We note that the values of $\sin^2\theta_{12}$, $\sin^2\theta_{13}$, $\sin^2\theta_{23}$ and $\delta$ predicted from the modular symmetry origin of texture zeros and quark-lepton unification (see Models \uppercase\expandafter{\romannumeral3} + $\mathcal{D}_{6}$ and \uppercase\expandafter{\romannumeral4} + $\mathcal{D}_{6}$ in the Ref.\cite{45}) are close to the ones shown in the special branch from  the correlation \uppercase\expandafter{\romannumeral1} in the top plots of Fig.\ref{fig:10}. In addition, the values of the mixing angles and the Dirac CP phase predicted from the $A_{5}^{'}$ modular group (see Lepton Models $\mathcal{L}_{134}$, $\mathcal{L}_{3}$, and $\mathcal{L}_{10}$ in the Ref.\cite{46}) also approach to the ones in the branch from the correlation \uppercase\expandafter{\romannumeral1} in the Fig.\ref{fig:10}. Thus, the geometric correlations may be realized in the lepton mixing models based on modular symmetries.

\section{The flavor ratio of high energy astronomical neutrinos}
In the previous section, we have studied the geometric correlations and the corresponding leptonic mixing parameters. Now we show the implications of geometric correlations on neutrino phenomenology. As a specific example, we discuss the impacts of geometric correlations on the flavor ratio of HANs at Earth in this section.

In recent years, a number of HANs events have been detected by the IceCube Collaboration in the energy range of TeV-PeV\cite{47,48,49,50}. These neutrinos travelled long distances of the cosmological scale. Their flavor transition probability $\overline{P}_{\alpha\beta}$ in the standard framework is expressed as
\begin{equation}
\label{eq:12}
  \overline{P}_{\alpha\beta}=|U_{\alpha1}|^{2}|U_{\beta1}|^{2}+|U_{\alpha2}|^{2}|U_{\beta2}|^{2}+|U_{\alpha3}|^{2}|U_{\beta3}|^{2},
\end{equation}
where $\alpha, \beta = e, \mu, \tau$, $U_{\alpha i}$ (i = 1, 2, 3) is the element of the PMNS matrix.
Using the flavor conversion matrix $\overline{P}$ and the flavor composition at the source of HANS, we can derive the flavor ratio at Earth, i.e., $\Phi^{E}=\overline{P}\Phi^{S}$.
Here $\Phi^{S}$ and $\Phi^{E}$ denote the flavor ratio at the source and that at Earth, respectively.
Since neither matter effects nor new physics effects are in favor of the neutron-decay as the sole source of the HANs\cite{51,52}, we take into account two typical sources in this paper, namely,
\begin{equation}
\label{eq:13}
\begin{array}{c}
muon-damping~~ source~~with ~~~ \Phi^{S}=(0, ~~1, ~~0)^T,\\
pion-decay ~~source~~with ~~~\Phi^{S}=(1/3, ~~2/3, ~~0)^T.
\end{array}
\end{equation}

\begin{figure}
  \centering
  \includegraphics[width=.49\textwidth]{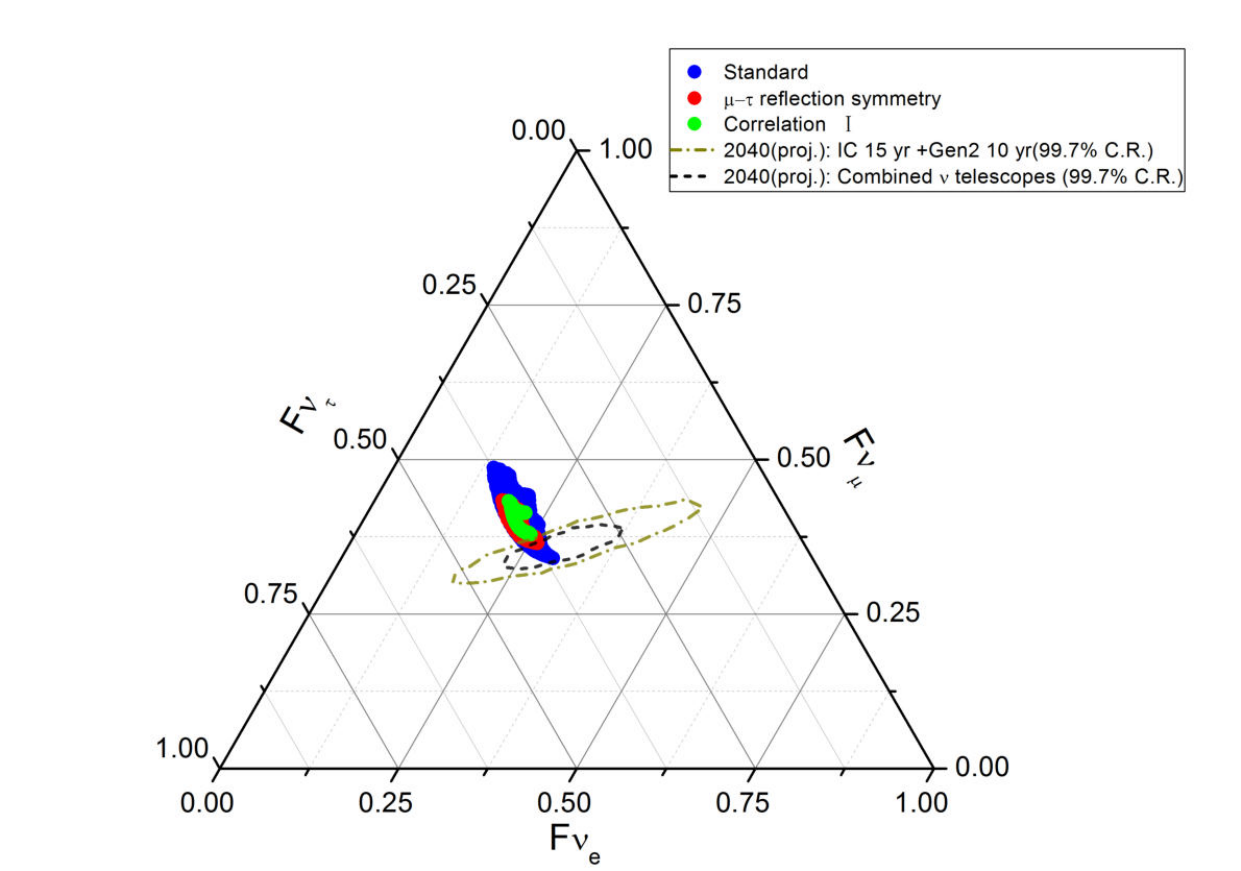}
  \hfill
  \includegraphics[width=.49\textwidth]{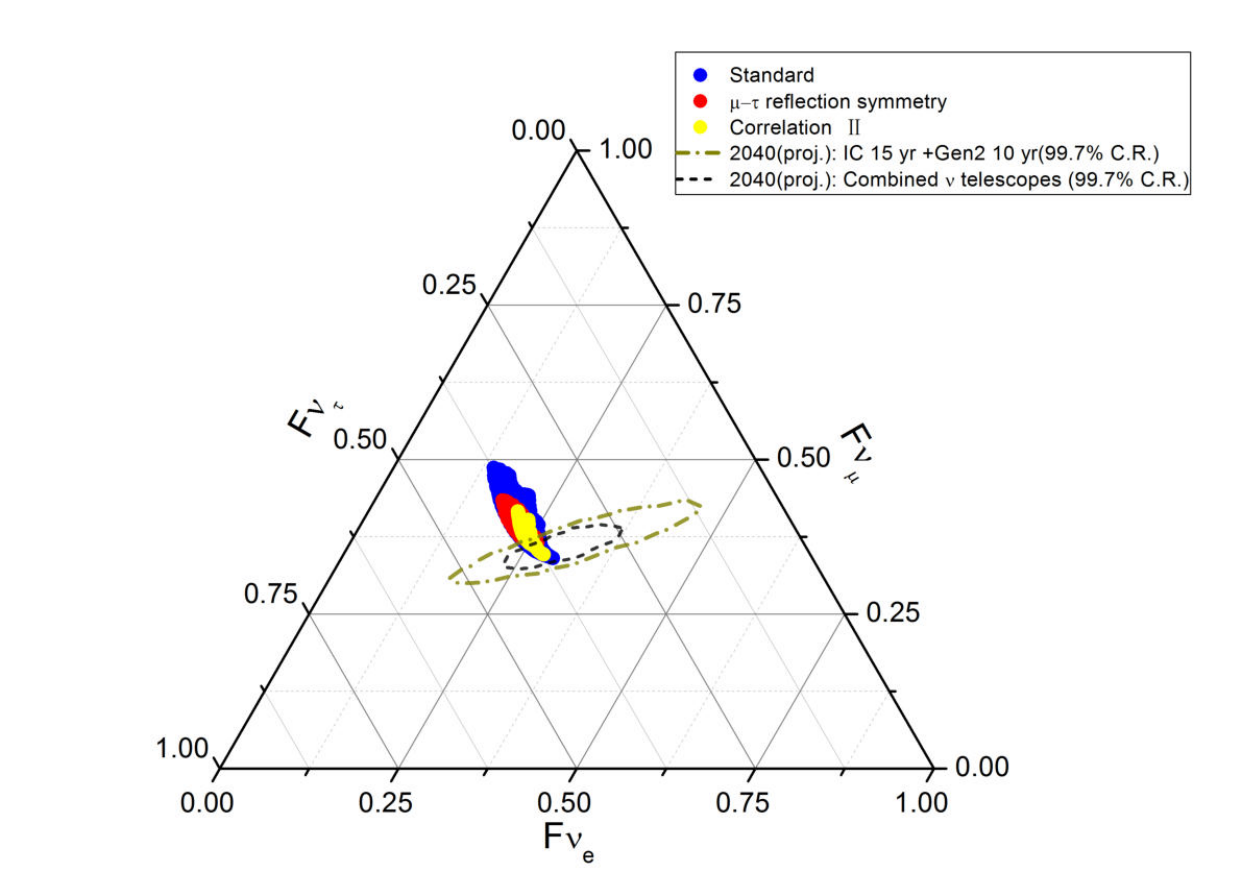}
  \hfill
  \includegraphics[width=.49\textwidth]{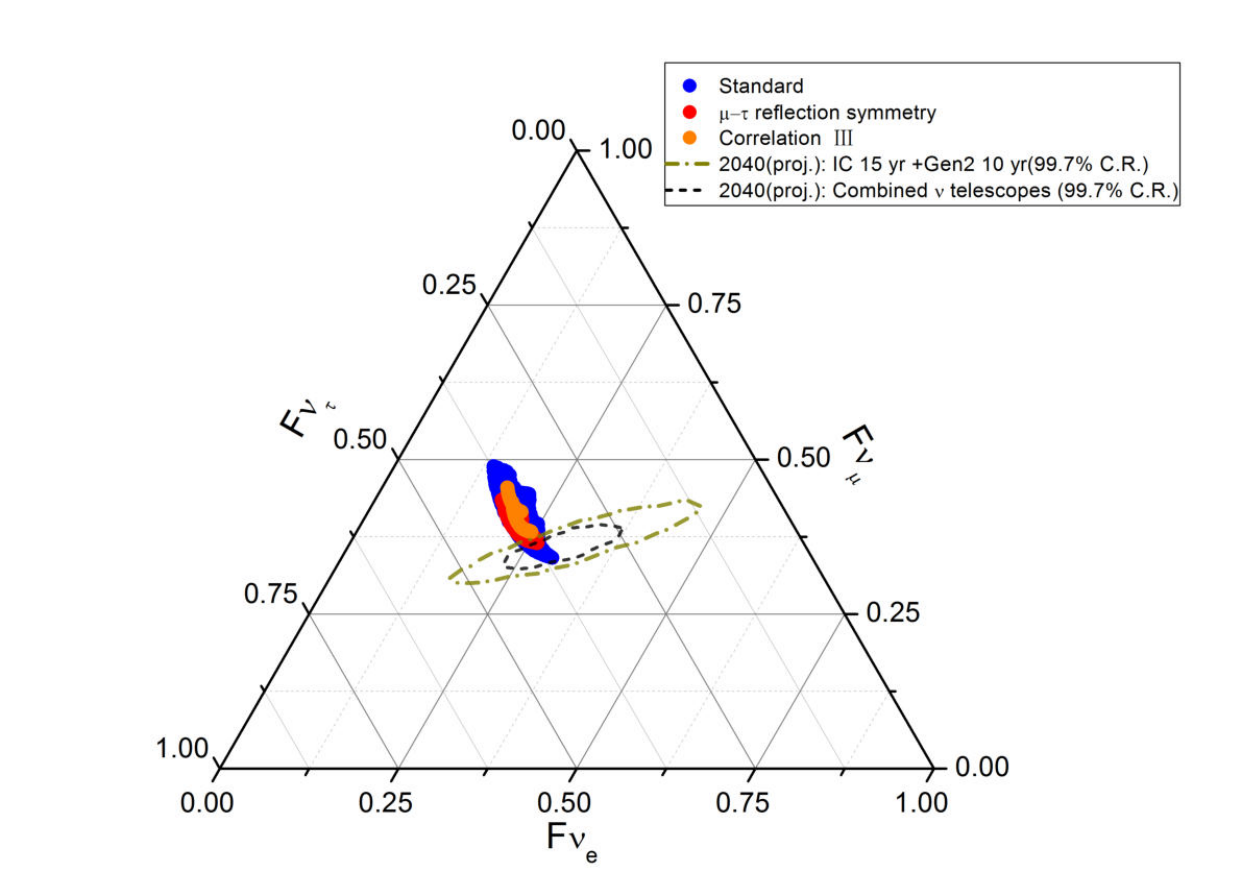}
  \hfill
  \includegraphics[width=.49\textwidth]{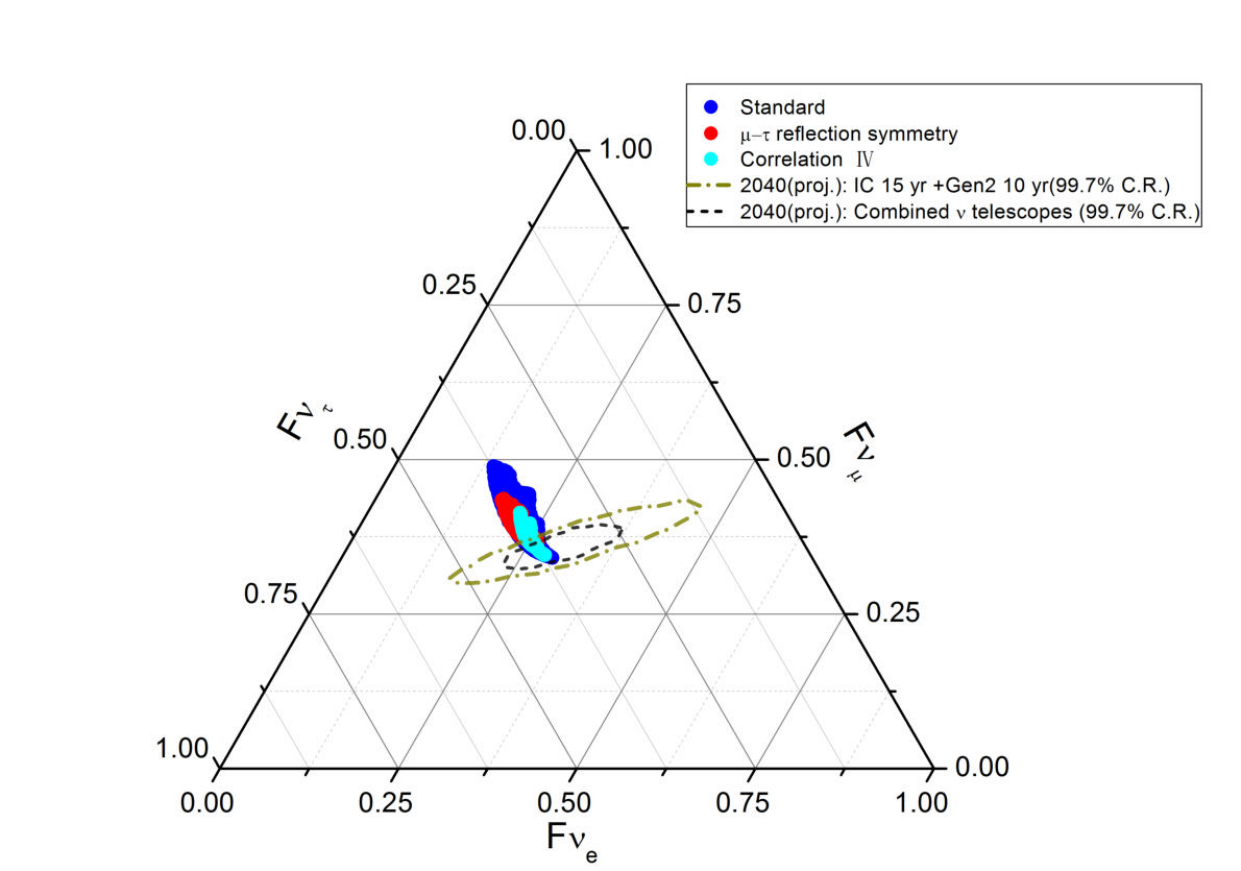}
  \caption{\label{fig:11} Ternary plot of the flavor ratio at Earth with the muon-damping decay source.  "Standard"  denotes the flavor ratio predicted with the leptonic mixing parameters at the $3\sigma$  level of the global fit data\cite{43}. "$\mu-\tau$ reflection symmetry" denotes the flavor composition that adds a constrain of $|\cos(\protect\overrightarrow{\mu}, \protect\overrightarrow{\tau})-1|<0.001$ to the "Standard" case. The geometric constraints of correlations \uppercase\expandafter{\romannumeral1} - \uppercase\expandafter{\romannumeral4} are defined in Tab.\ref{tab:5} and Eq.\ref{eq:11}. The 2040 $3\sigma$ credible region
with the pion-decay source based on IceCube and IceCube-Gen2 is taken from Ref.\cite{53}. The $3\sigma$ credible region with the pion-decay source based on all TeV-PeV neutrino telescopes available in 2040 is taken from Ref.\cite{54}.}
\end{figure}

For a given source, the uncertainty of the predicted flavor ratio at Earth is large due to the imprecise leptonic mixing parameters. As is known, the constraints of geometric correlations can significantly reduce the ranges of the leptonic mixing parameters. Accordingly, the precision of $\Phi^{E}$ can be notably improved. Here we show the impacts of the representative geometric correlations on $\Phi^{E}$ in Fig.\ref{fig:11} - Fig.\ref{eq:13}.
For the sake of comparison, the flavor ratio predicted with the leptonic mixing parameters at the  $3\sigma$  level of the global fit data\cite{43} and  those constrained by the $\mu-\tau$ reflection symmetry are
also shown in the ternary plots.
\begin{figure}
  \centering
  \includegraphics[width=.49\textwidth]{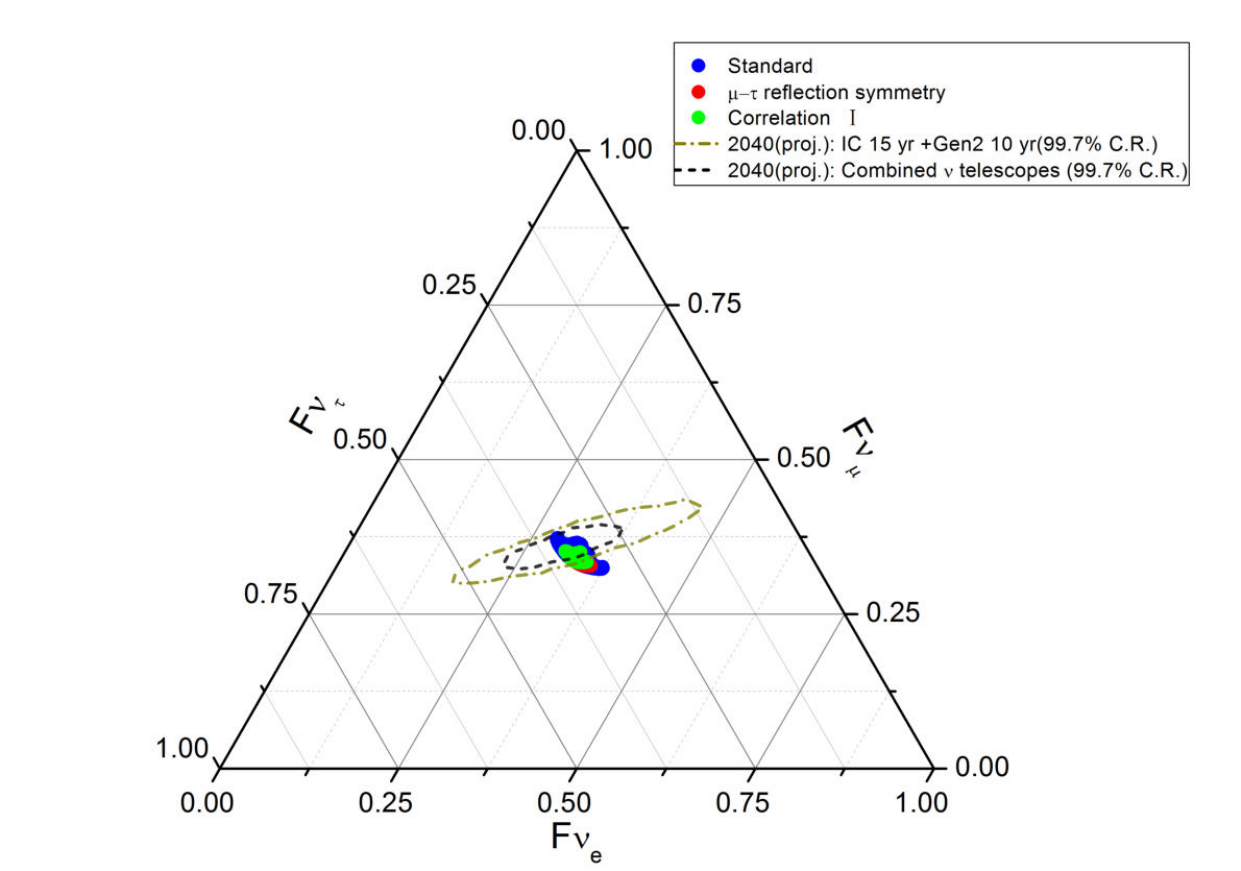}
  \hfill
  \includegraphics[width=.49\textwidth]{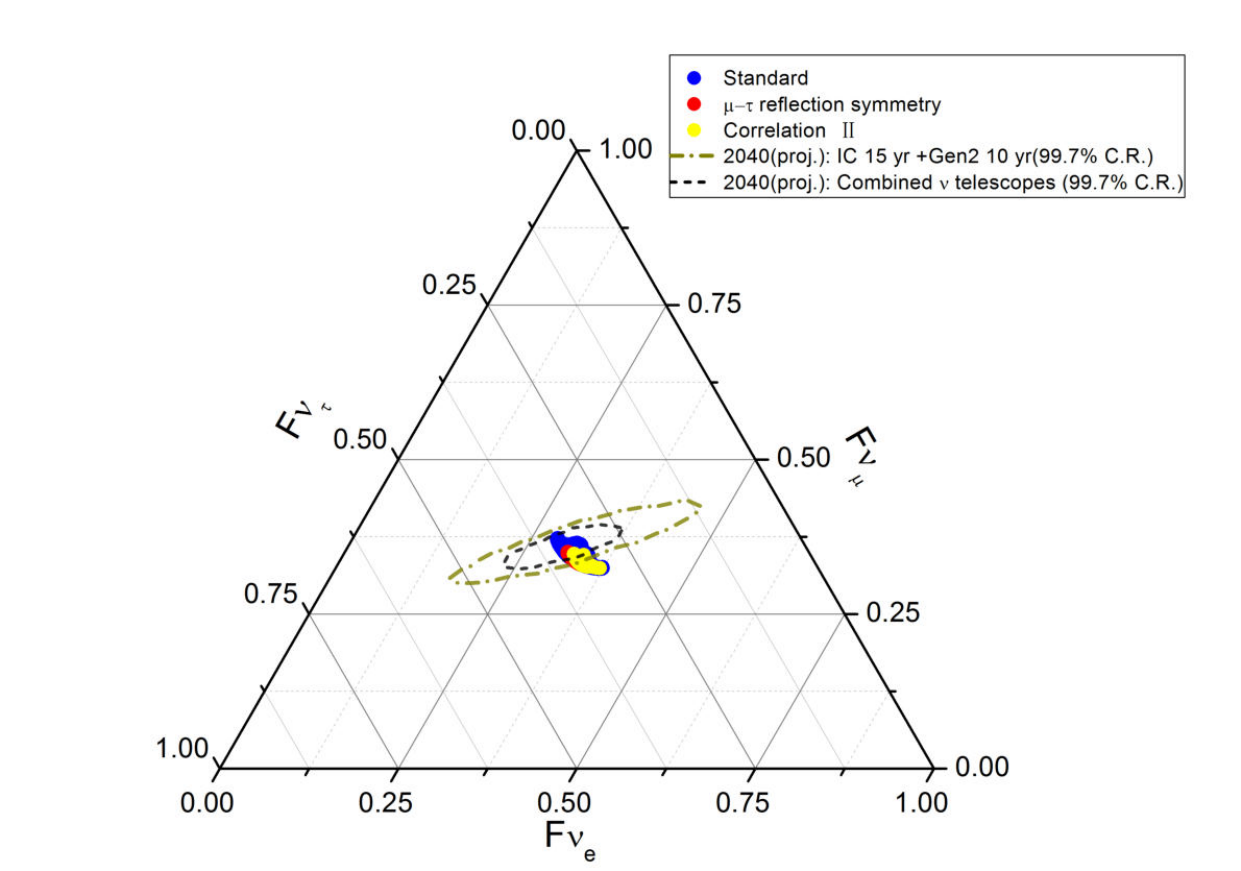}
  \hfill
  \includegraphics[width=.49\textwidth]{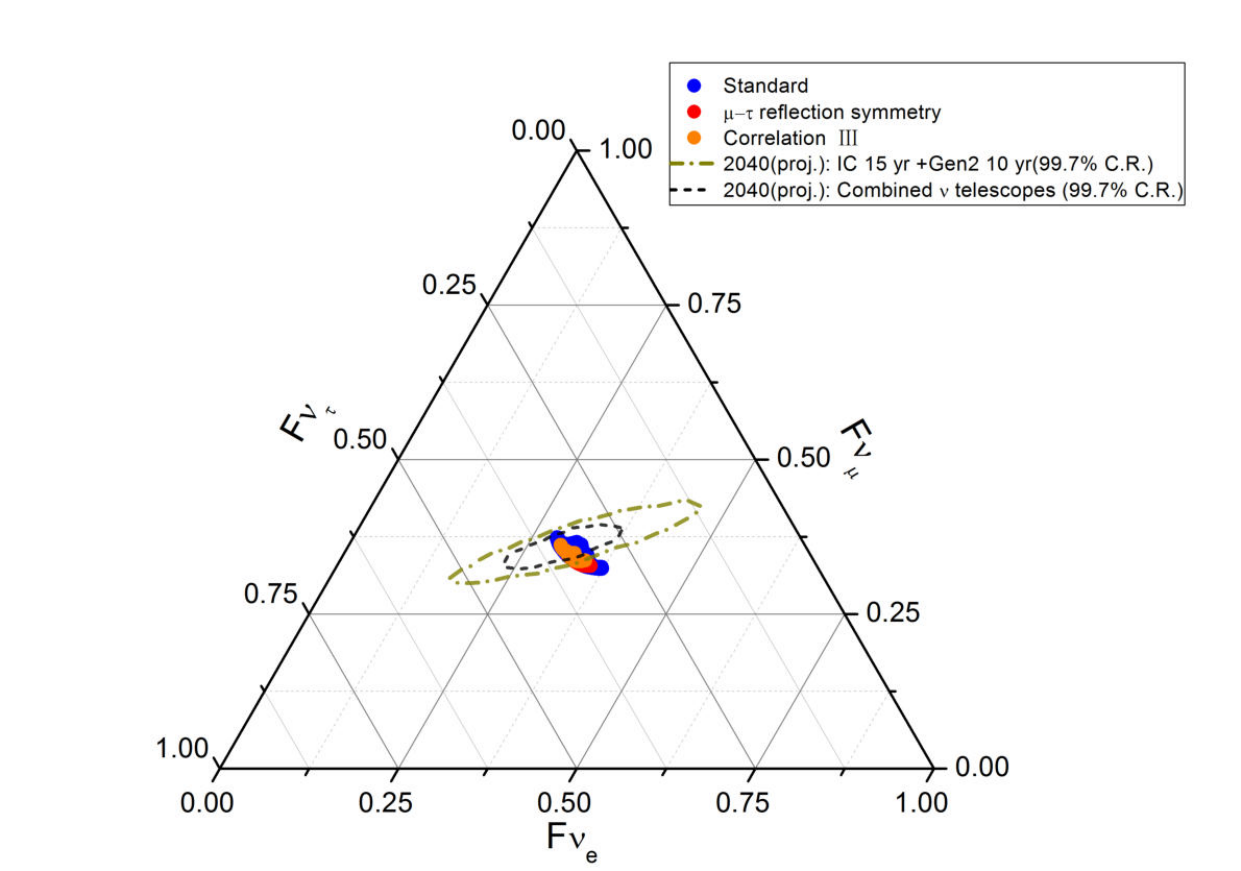}
  \hfill
  \includegraphics[width=.49\textwidth]{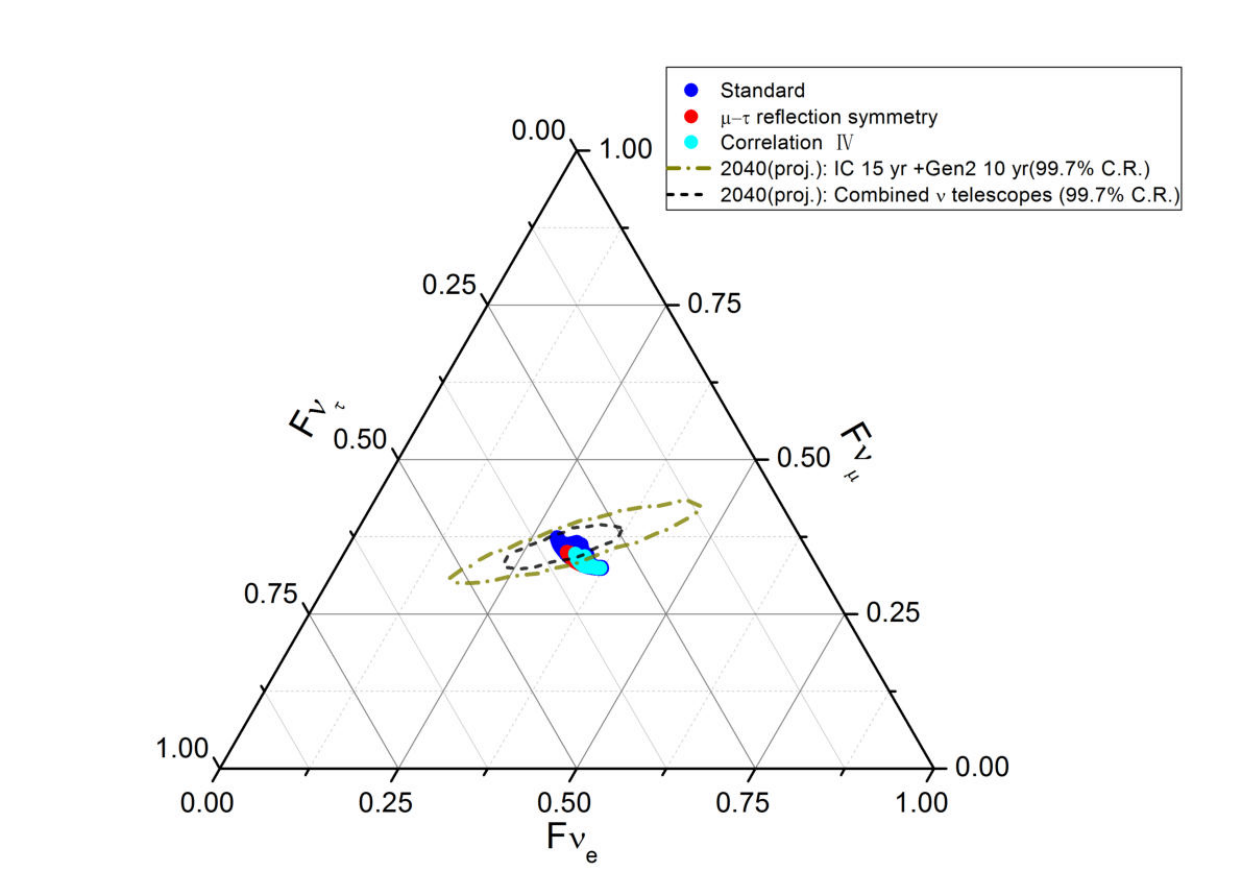}
  \caption{\label{fig:12} Ternary plot of the flavor ratio at Earth with the pion-decay source based on the geometric constraints of the correlations \uppercase\expandafter{\romannumeral1} - \uppercase\expandafter{\romannumeral4}. The conventions of parameters and colors are the same as those in Fig.\ref{fig:11}.}
\end{figure}
\begin{figure}
  \centering
  \includegraphics[width=.49\textwidth]{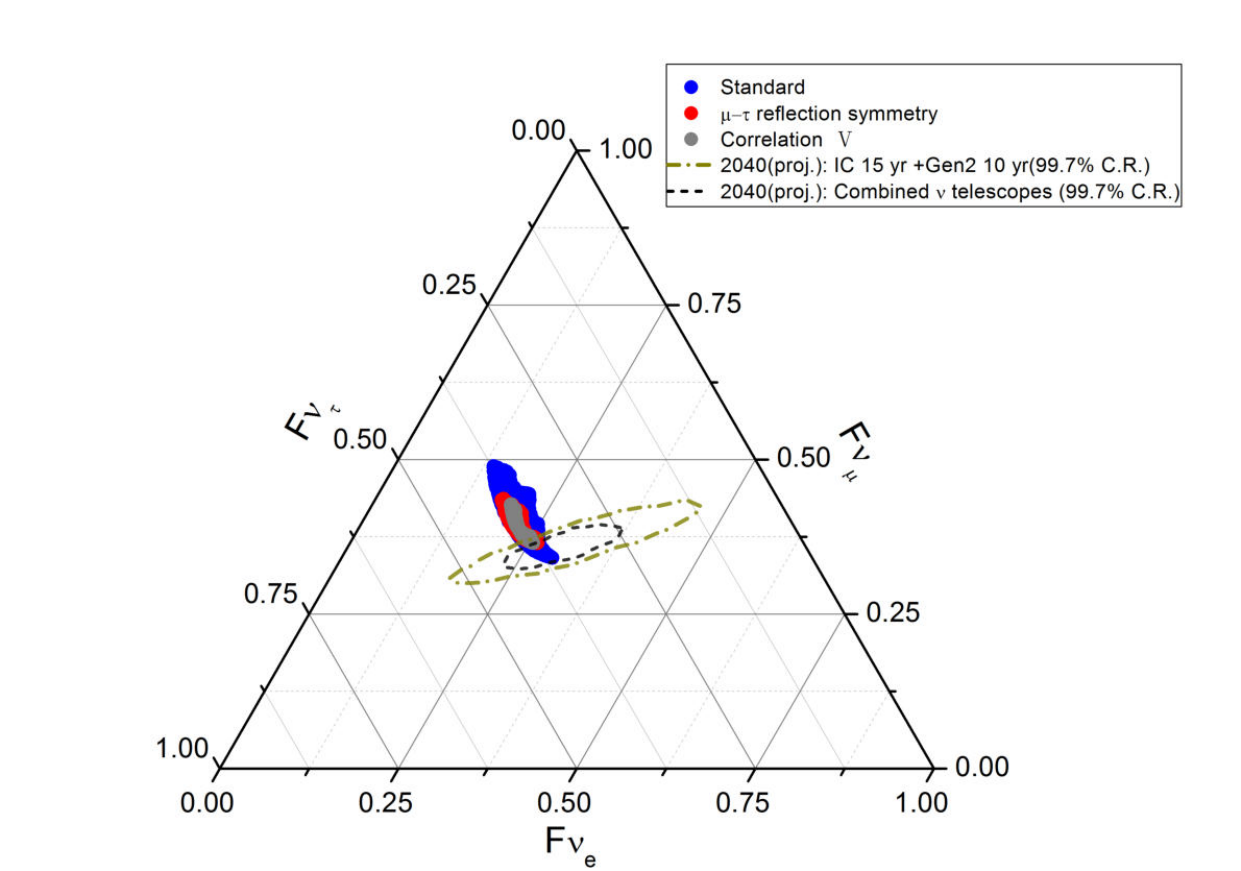}
  \hfill
  \includegraphics[width=.49\textwidth]{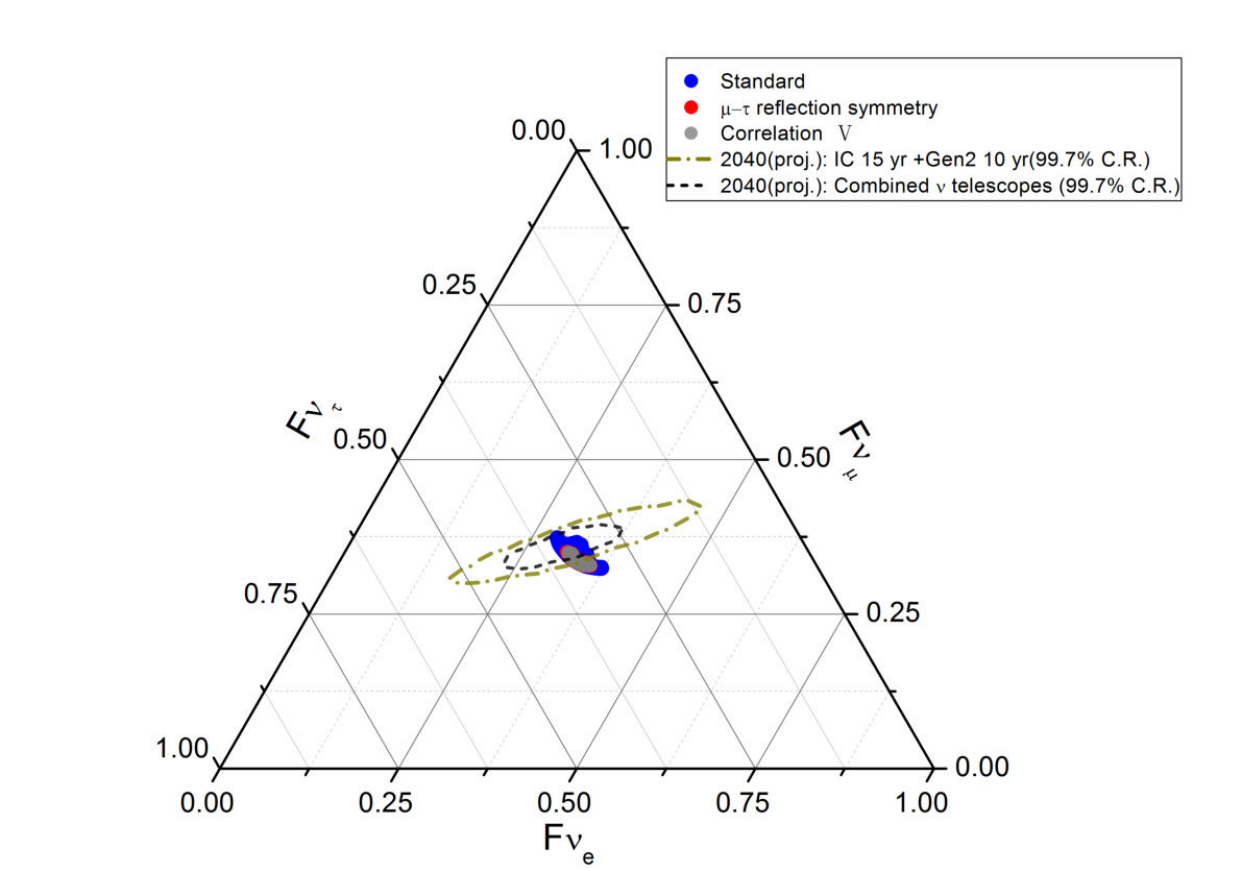}
  \caption{\label{fig:13} Ternary plot of the flavor ratio at Earth with the muon-damping decay source (left panel) and the pion-decay source (right panel) based on the geometric constraints of the correlations \uppercase\expandafter{\romannumeral5}. The conventions of parameters and colors are the same as those in Fig.\ref{fig:11}.}
\end{figure}
From these figures, we obtain the following observations:\\
(a) For both sources of HANs, the uncertainty of the predicted  flavor ratio at  Earth  is obviously decreased by the geometrical constraints of every correlation.
In contrast to the strict $\mu-\tau$ reflection symmetry, the geometrical correlations have more notable impacts on the predicted flavor ratio .\\
(b) For the muon-damping source (see Fig.\ref{fig:11} and the left panel of Fig.\ref{fig:13}), the predicted flavor ratio constrained by the correlation \uppercase\expandafter{\romannumeral1} and/or \uppercase\expandafter{\romannumeral3} is outside the 3$\sigma$ credible region with the pion-decay source based on all neutrino telescopes\cite{54}.
However, a small part of the flavor ratio at Earth, which constrained by the correlation \uppercase\expandafter{\romannumeral2}, \uppercase\expandafter{\romannumeral4} and/or \uppercase\expandafter{\romannumeral5} falls into the 3$\sigma$ credible region with the pion-decay source.
Therefore, with the help of geometrical correlations, the muon-damping decay source may be discriminated from the pion-decay source in the near future.\\
(c) For the pion-decay source (see Fig.\ref{fig:12} and the right panel of Fig.\ref{fig:13}), the primary part of the flavor ratio constrained by the correlation \uppercase\expandafter{\romannumeral2} and/or \uppercase\expandafter{\romannumeral4} is outside the 2040 3$\sigma$ credible region. Thus, the correlations \uppercase\expandafter{\romannumeral2} and \uppercase\expandafter{\romannumeral4} with this source would be stringently constrained by the observations of the neutrino telescopes in the future.

\section{Conclusions}
The $\mu-\tau$ reflection symmetry has been proposed for twenty years. Its prediction still satisfies the constraints of the recent global fit data of neutrino oscillations.
From the geometric perspective, this symmetry and its breaking can be represented  by the angles between the row vectors of the magnitude of the leptonic mixing matrix.
Employing the geometric quantities, we studied the promising correlations between the leptonic mixing parameters.
On the bases of the mixing parameters constrained by the geometric correlations,
the flavor ratio of HANs at Earth with typical sources has been discussed. Our main results are summarized as follows.

First, the data points of the included angles of the row vectors are concentrated on several special regions in the scattered plots.
In the IO case, one of the regions corresponds to the approximated $\mu-\tau$ reflection symmetry.
From other regions, promising correlations of leptonic mixing parameters are read out.

Second, we find that $\sin^2\theta_{12}$ is sensitive to the correlation ($\cos(\overrightarrow{e_{0}},\overrightarrow{\mu_{0}}), \cos(\overrightarrow{e_{0}},\overrightarrow{\tau_{0}})$) in both the NO case and IO case.
In contrast, $\sin^2\theta_{23}$ and $\delta$ changes notably under the moderate variation of the correlation ($\cos(\overrightarrow{\mu_{0}},\overrightarrow{\tau_{0}}), \cos(\overrightarrow{e_{0}},\overrightarrow{\mu_{0}})$) and/or ($\cos(\overrightarrow{\mu_{0}},\overrightarrow{\tau_{0}}), \cos(\overrightarrow{e_{0}},\overrightarrow{\tau_{0}})$).
Furthermore, the ranges of $\sin^2\theta_{23}$ and $\delta$ display disconnected branches. Thus, the included angles of the row vectors can serve as sensitive indexes of the leptonic mixing pattern to test by the neutrino oscillation experiments in future.

Third, when the correlation of the three included angles  ($\cos(\overrightarrow{e_{0}},\overrightarrow{\mu_{0}}), \cos(\overrightarrow{e_{0}},\overrightarrow{\tau_{0}}), \cos(\overrightarrow{\mu_{0}},\overrightarrow{\tau_{0}}$)) is given, $\sin^2\theta_{12}$ is linearly dependent on $\sin^2\theta_{13}$, and
$\sin^2\theta_{23}$ and $\delta$ can take discrete values. Hence, the uncertainties of the leptonic mixing are remarkably decreased by the geometric correlations.
Correspondingly, the prediction of the flavor ratio of HANs at Earth with typical sources is notably improved. With the help of the  special correlations, the discrimination of the sources may be available in the near future.

\vspace{0.08cm}

\acknowledgments
 This work is supported by the National Natural Science Foundation of China under grant No. 12065007, the Guangxi Scientific Programm Foundation under grant No. Guike AD19110045, the Research Foundation of Guilin University of Technology under grant No. GUTQDJJ2018103. \\

\bibliography{refs}

\end{document}